\documentclass[12pt]{article}
\usepackage{amsfonts}
\usepackage{amssymb}%,letterspace}
\usepackage{graphics,psboxit,amsmath}
\usepackage{subfigure}
\usepackage{graphicx}
\usepackage{verbatim}
\usepackage{epic}

\def\hybrid{\topmargin 0pt \oddsidemargin 0pt %%%%%%%%%%%%%% Archive-30pt
        \headheight 0pt \headsep 0pt
        \textwidth 16,5cm % A4 paper
        \textheight 23,5cm % A4 paper
        \marginparwidth .875in
        \parskip 5pt plus 1pt \jot = 1.5ex}

\hybrid

\def\baselinestretch{1.2}

\catcode`\@=11

\def\marginnote#1{}
\newcount\hour
\newcount\minute
\newtoks\amorpm
\hour=\time\divide\hour by60
\minute=\time{\multiply\hour by60 \global\advance\minute by-\hour}
\edef\standardtime{{\ifnum\hour<12 \global\amorpm={am}%
        \else\global\amorpm={pm}\advance\hour by-12 \fi
        \ifnum\hour=0 \hour=12 \fi
        \number\hour:\ifnum\minute<10 0\fi\number\minute\the\amorpm}}
\edef\militarytime{\number\hour:\ifnum\minute<10 0\fi\number\minute}
\def\draftlabel#1{{\@bsphack\if@filesw {\let\thepage\relax
   \xdef\@gtempa{\write\@auxout{\string
      \newlabel{#1}{{\@currentlabel}{\thepage}}}}}\@gtempa
   \if@nobreak \ifvmode\nobreak\fi\fi\fi\@esphack}
        \gdef\@eqnlabel{#1}}
\def\@eqnlabel{}
\def\@vacuum{}
\def\draftmarginnote#1{\marginpar{\raggedright\scriptsize\tt#1}}

\def\draft{\oddsidemargin -.5truein
        \def\@oddfoot{\sl preliminary draft \hfil
        \rm\thepage\hfil\sl\today\quad\militarytime}
        \let\@evenfoot\@oddfoot \overfullrule 3pt
        \let\label=\draftlabel
        \let\marginnote=\draftmarginnote
   \def\@eqnnum{(\theequation)\rlap{\kern\marginparsep\tt\@eqnlabel}%
\global\let\@eqnlabel\@vacuum} }

\def\draft2{
        \def\@oddfoot{\sl preliminary draft \hfil
        \rm\thepage\hfil\sl\today\quad\militarytime}
        \let\@evenfoot\@oddfoot \overfullrule 3pt
        \let\label=\draftlabel
        \let\marginnote=\draftmarginnote
   \def\@eqnnum{(\theequation)\rlap{\kern\marginparsep\tt\@eqnlabel}%
\global\let\@eqnlabel\@vacuum} }

\def\preprint{\twocolumn\sloppy\flushbottom\parindent 2em
        \leftmargini 2em\leftmarginv .5em\leftmarginvi .5em
        \oddsidemargin -.5in \evensidemargin -.5in
        \columnsep .4in \footheight 0pt
        \textwidth 10.in \topmargin -.4in
        \headheight 12pt \topskip .4in
        \textheight 6.9in \footskip 0pt
        \def\@oddhead{\thepage\hfil\addtocounter{page}{1}\thepage}
        \let\@evenhead\@oddhead \def\@oddfoot{} \def\@evenfoot{} }

\def\numberbysection{\@addtoreset{equation}{section}
        \def\theequation{\thesection.\arabic{equation}}}

\def\underline#1{\relax\ifmmode\@@underline#1\else
        $\@@underline{\hbox{#1}}$\relax\fi}

\def\titlepage{\@restonecolfalse\if@twocolumn\@restonecoltrue\onecolumn
     \else \newpage \fi \thispagestyle{empty}\c@page\z@
        \def\thefootnote{\fnsymbol{footnote}} }

\def\endtitlepage{\if@restonecol\twocolumn \else \newpage \fi
        \def\thefootnote{\arabic{footnote}}
        \setcounter{footnote}{0}} %\c@footnote\z@ }

\catcode`@=12
\relax

\def\figcap{\section*{Figure Captions\markboth
        {FIGURECAPTIONS}{FIGURECAPTIONS}}\list
        {Figure \arabic{enumi}:\hfill}{\settowidth\labelwidth{Figure
999:}
        \leftmargin\labelwidth
        \advance\leftmargin\labelsep\usecounter{enumi}}}
 \relax
\def\tablecap{\section*{Table Captions\markboth
        {TABLECAPTIONS}{TABLECAPTIONS}}\list
        {Table \arabic{enumi}:\hfill}{\settowidth\labelwidth{Table
999:}
        \leftmargin\labelwidth
        \advance\leftmargin\labelsep\usecounter{enumi}}}
 \relax
\def\reflist{\section*{References\markboth
        {REFLIST}{REFLIST}}\list
        {[\arabic{enumi}]\hfill}{\settowidth\labelwidth{[999]}
        \leftmargin\labelwidth
        \advance\leftmargin\labelsep\usecounter{enumi}}}
 \relax
\makeatletter
\newcounter{pubctr}
\def\publist{\@ifnextchar[{\@publist}{\@@publist}}
\def\@publist[#1]{\list
        {[\arabic{pubctr}]\hfill}{\settowidth\labelwidth{[999]}
        \leftmargin\labelwidth
        \advance\leftmargin\labelsep
        \@nmbrlisttrue\def\@listctr{pubctr}
        \setcounter{pubctr}{#1}\addtocounter{pubctr}{-1}}}
\def\@@publist{\list
        {[\arabic{pubctr}]\hfill}{\settowidth\labelwidth{[999]}
        \leftmargin\labelwidth
        \advance\leftmargin\labelsep
        \@nmbrlisttrue\def\@listctr{pubctr}}}
 \relax
\makeatother

\def\ba{\begin{equation}}
\def\ea{\end{equation}}

\def\del{\partial}

\def\d{\delta}

\def\l{\lambda}

\def\no{\noindent}

\def\qq{\qquad}

\def\IR{\relax{\rm I\kern-.18em R}}

\begin{document}
%\draft2

%\renewcommand{\theequation}{\arabic{equation}}
%\renewcommand{\theequation}{\thesection.\arabic{equation}}

\renewcommand{\theequation}{\thesection.\arabic{equation}}
\csname @addtoreset\endcsname{equation}{section}

\newcommand{\eqn}[1]{(\ref{#1})}
\newcommand{\be}{\begin{eqnarray}}
\newcommand{\ee}{\end{eqnarray}}
\newcommand{\non}{\nonumber}

\begin{titlepage}
\strut\hfill
\vskip 1.3cm
\begin{center}

{\large \bf Selected Topics in Classical Integrability}

\vskip 0.5in

{\bf Anastasia Doikou}\footnote{\tt This article is based on a series of lectures presented
at the University of Oldenburg in July 2011.}
\vskip 0.1in

{\footnotesize
Department of Engineering Sciences, University of Patras,\\
GR-26500 Patras, Greece}\\

\vskip .1in

%\vskip -.15in

{\footnotesize {\tt E-mail: adoikou$@$upatras.gr}}\\

\end{center}

%\vfill
\vskip .6in

\centerline{\bf Abstract}

Basic notions regarding classical integrable systems are reviewed. An algebraic description of the classical integrable models together with the zero curvature condition description is presented. The classical $r$-matrix approach for discrete and continuum classical integrable models is introduced. Using this framework the associated classical integrals of motion and the corresponding Lax pair are extracted based on algebraic considerations. Our attention is restricted to classical discrete and continuum integrable systems with periodic boundary conditions. Typical examples of discrete (Toda chain, discrete NLS model) and continuum integrable models (NLS, sine-Gordon models and affine Toda field theories) are also discussed.

\no

\vfill
\no

\end{titlepage}
\vfill
%\vskip .5cm
%\noindent

%\end{titlepage}
%\vfill
\eject

%\def\baselinestretch{1.2}
%\baselineskip 10 pt
%\noindent

\tableofcontents

\def\baselinestretch{1.2}
\baselineskip 20 pt
\no

\section{Introduction}

This is a pedagogical review aiming at introducing basic notions regarding classical integrable models. It is based on a
series of Lectures given at the University of Oldenburg. These lectures are mainly addressed to graduate students or other researchers
who wish to acquire some fundamental knowledge on certain topics within classical integrability \cite{ftbook}--\cite{ZSh}.

The notion of complete integrability goes back to the celebrated Liouville's theorem for a system with a finite number of degrees
of  freedom \cite{arnold}. In 70's the notion of integrability was extended to models with infinite degrees of freedom, that is 1+1 integrable
field theories with generic non-linear interactions; such typical examples, which will be reviewed here are the non-linear Schrodinger
model (NLS), the sine-Gordon model and its generalizations, the affine Toda field theories (ATFT), among others (see e.g. \cite{ftbook, BBT} and references therein). The idea of quantum
integrability is relevant to the so called quantum inverse scattering method, an elegant algebraic technique mainly developed by the St. Petersburg group (see e.g. \cite{FTS}--\cite{tak}, for a recent review see \cite{doikouint} and references therein). Perhaps the greatest
appeal of integrability is that it offers an exact framework for the investigation of a wide range of physical systems without resort to perturbative methods, which are the most common techniques when studying any physical system. In addition to the significance
of these models per se, and their rich mathematical structure there are various immediate applications and connections with other
related areas of research, such as string and gauge theories, condensed matter physics, gravity etc.

The main focus in this presentation lies on the description of classical discrete and continuum integrable systems via the Hamiltonian formalism based on the classical $r$-matrix approach \cite{skl, sts}. More precisely, the outline of this article is as follows: in the next Section some preliminary notions such as the Poisson
structure are introduced, and Liouville's theorem is briefly reviewed.
In Section 3 linear Poisson algebraic structures are introduced and integrable models ruled by linear algebras, such as the Toda chain are described. In particular, the first integrals on motion are extracted and the
time components of the corresponding Lax pairs are explicitly constructed from first principles. In section 4 we introduce the notion of quadratic Poisson structure, and through this notion we construct
discrete one-dimensional discrete integrable models. An explicit construction of the conserved quantities and the associate Lax pairs is presented as well. In the next Section typical examples of physical systems ruled by quadratic classical algebras, such as the discrete NLS model as well as the Toda chain (via the ``dual'' quadratic description due to Sklyanin), are presented. In Section 6 we describe
classical continuum integrable models in the spirit presented in the previous sections, based again on quadratic Poisson structures.
Typical examples such as the NLS and sine-Gordon models, affine Toda field theories (ATFT) and the Liouville model are reviewed in
Section 7. In the last section a systematic prescription for obtaining suitable continuum limits of discrete integrable models in such a way that integrability is preserved is outlined. A considerable number of relevant exercises are provided for the interested reader throughout the text.

\section{Preliminaries}

The notion of Poisson bracket, which is used to define a Poisson algebra (see e.g. \cite{ftbook, BBT, arnold} and references therein) is briefly reviewed.
Consider a phase space ${\mathfrak G}_{2n}$ endowed with Poisson structure $\{ \ ,\ \}$ and local canonical variables $q_i,\ p_i$. Given any two functions $f(\{q_i\},\ \{p_i\}),\ g(\{q_i\},\ \{p_i\})$ of the canonical
variables associated to the Poisson bracket takes by definition the form:
\be
\Big \{f,\ g \Big \} = \sum_{i=1}^n \Big ({\partial f \over \partial p_i} {\partial g\over q_i} - {\partial f \over
\partial q_i} {\partial g \over \partial p_i}\Big )
\ee
and clearly
\be
\Big \{q_i,\ q_j \Big \} = \Big \{ p_i,\ p_j\Big \}=0, ~~~~~\Big \{p_i,\ q_j \Big \} = \delta_{ij}.
\ee

For any three functions $f,\ g,\ h$ of the canonical variables the so called Jacobi identity holds, that is:
\be
\Big \{f,\ \Big \{g,\ h\Big \}\Big \} +\Big \{g,\ \Big \{h,\ f\Big \}\Big \} +\Big \{h,\ \Big \{f,\ g\Big \}\Big \} =0.
\ee
Also, Leibniz's law, known as Poisson property applies:
\be
\Big \{f, g h\Big \} = g \Big \{f,\ h\Big \} + \Big \{f,\ g \Big \} h.
\ee

Hamilton's equations of motion may be derived in terms of the Poisson brackets. Let a function $f(\{q_i\},\ \{p_i\},\ t)$, then
\be
{d f \over dt} = \sum_i^n{\partial f \over q_i} \dot q_i +  \sum_i^n {\partial f \over p_i} \dot p_i + {\partial f \over \partial t}.
\ee
But
\be
\dot q_i = {\partial H \over \partial p_i}, ~~~~~\dot p_i = - {\partial H \over \partial q_i}
\ee
(the dot denotes derivative with respect to time) therefore we conclude:
\be
{d f \over dt} =\Big  \{H,\ f \Big  \}  + {\partial f \over \partial t}.
\ee
For $f$ being a constant of motion one then has
\be
\Big \{f,\ H \Big \} + {\partial f \over \partial t} = 0.
\ee

In order for a system to be integrable all the integrals of motion should be in mutual involution. i.e. to Poisson commute.
More precisely, let us recall the celebrated Liouville's theorem (see e.g. \cite{BBT, arnold}): consider a system with
 $n$ degrees of freedom in the phase space ${\mathfrak G}_{2n}$ with coordinates $q_1, \dots, q_n,\ p_1 \ldots, p_n$,
 let also $H(\{q_i\},\ \{p_i\})$ be the Hamiltonian of the system. The system is completely integrable if there exist $n$
 independent functions $I_i(\{q_i\},\ \{p_i\})$ defined on the phase space such that:
\be
\Big \{ H,\ I_i \Big \} =0, ~~~~~\Big \{ I_i,\ I_j \Big \} =0,
\ee
$I_i$ are in fact charges in involution and they are known as integrals of motions or conserved quantities given that they satisfy:
 \be
{d I_j \over dt} = \Big \{H,\ I_j \Big \} =0.
\ee
In this case it is possible to introduce a complementary set of functions $\varphi_i(\{q_j\},\ \{p_j\} )$, $i,\ j \ \in \{1,\ \ldots, n\}$ \cite{arnold}, and we may then describe the change of coordinates on the phase space:
\be
\Big (q_i,\ p_i \Big )\ \mapsto \ \Big (I_i,\ \varphi_i \Big ).
\ee
The new coordinates $\Big (I_i,\ \varphi_i \Big )$ define the so-called {\it action-angle} variables, and the equations
of motion become:
\be
&&{d I_j \over dt} = \Big \{H,\ I_j \Big \} =0 \ \Rightarrow \ I_j(t) = \mbox{constant}\non\\
&& {d \varphi_j \over dt}  =  \Big \{H,\ \varphi_j \Big \} = \omega_j \ \Rightarrow \ \varphi_j(t) = \omega_j\ t + \varphi_j(0)
\ee
for certain functions $\omega_j(\{I_i\})$. The time evolution of an integrable system is linear if we choose a suitable set of coordinates (angle-action). Although the existence of $\Big (I_i,\ \varphi_i \Big )$ is guaranteed from Liouville's theorem (see e.g. \cite{arnold}) their explicit construction is an intriguing task depending on the specifics of the integrable system under consideration

Having introduced the basic notions on the Poisson structures we are now ready to proceed with the description of the
algebraic setting underlying classical discrete and continuous integrable systems.

\section{Linear Poisson structure}
We start our review considering integrable models that obey linear algebraic Poisson structures. A typical example
in this category is the Toda chain. We first introduce the Lax pair \cite{lax}
associated to such a system. It is a pair of elements of a Lie algebra ${\cal G}$, $(L,\ {\mathbb A})$ satisfying the so called zero curvature condition:
\be
{d L \over d t} = \Big [L,\ {\mathbb A} \Big ]. \label{zero}
\ee
Assuming ${\cal G}$ is a loop algebra ${\cal G} \otimes {\mathbb C}(\lambda)$, the associated spectral problem is then expressed as
\be
&&L(\lambda)\ \psi = u\ \psi \non\\
&&\det \Big (L(\lambda) -u \Big ) =0.
\ee
The integrals of motion associated to such a system are then naturally provided by the traces of powers of $L$ i.e. (see also
\cite{BBT, avandoikou} and references therein),
\be
t^{(n)}(\lambda) = tr\ (L^n(\lambda)). \label{tnn}
\ee
Expansion in powers of $\lambda$ (or $1\over \lambda$) is expected to provide all required integrals of motion of the system under consideration.

It is quite straightforward to show that the latter provide a family of mutually Poisson quantities as long as $L$
satisfies the following linear algebraic relations:
\be
\Big \{L_a(\lambda),\ L_b(\mu) \Big \}= \Big [r_{ab}(\lambda - \mu),\ L_a(\lambda)+L_b(\mu) \Big ] \label{linear}
\ee
$L$ is in general an $n \times n$ matrix with entries being fields, representations of the underlying classical
algebra, and  $r$ is an $n^2 \times n^2$ matrix which satisfies the classical Yang-Baxter equation \cite{sts}
\be
\Big [r_{12}(\lambda_1 -\lambda_2),\ r_{13}(\lambda_1)+r_{23}(\lambda_2) \Big ] + \Big [r_{13}(\lambda_1),\  r_{23}(\lambda_2)\Big ]=0.
\label{cybe} \ee
The $r$-matrix acts on the tensor product $V \otimes V$. In general the notation that will be used henceforth is:
$A_1 =  A \otimes {\mathbb I}$, $~B_2 = {\mathbb I} \otimes B$ and $A_1\ B_2 = A \otimes B$. Moreover, it is clear that the classical
Yang-Baxter equation acts on ${\mathbb V} \otimes {\mathbb V} \otimes {\mathbb V}$ and  $r_{12} = r \otimes {\mathbb I}$ i.e.
the $r$ matrix acts non trivially on the first two spaces, $r_{23} = {\mathbb I} \otimes r$ and so on.
\\
\\
\textbf{\underline{Proposition 3.1.}}
Based on the linear relations (\ref{linear}) show that:
\be
\Big \{ t^{(n)}(\lambda),\ t^{(m)}(\mu)\Big \} =0.
\ee
\textit{\underline{Proof}}: We recall the definition of $t^{(n)}$ and substitute:
\be
tr_{ab} \Big \{L_a^n(\lambda),\ L_b^m(\mu) \Big\}= n m\ tr_{ab} \Big ( L_{a}^{n-1}(\lambda) \Big \{ L_a(\lambda),\
L_b(\mu)\Big\}L_b^{m-1}(\mu) \Big ) = \ldots
\ee
The latter expression becomes after recalling (\ref{linear})
\be
\ldots &=& n m\ tr_{ab} \Big ( L_a^{n-1}(\lambda) \Big [ r_{ab}(\lambda -\mu),\ L_a(\lambda) +L_b(\mu)\Big ]L_b^{m-1}(\mu)
\Big )  \non\\
&=& n m \ tr_{ab} \Big ( L_a^{n-1}(\lambda)L_b^{m-1}(\mu) r_{ab}(\lambda-\mu) (L_a(\lambda) + L_b(\mu)) \Big )\non\\
&-&  n m \ tr_{ab} \Big ( L_a^{n-1}(\lambda)L_b^{m-1}(\mu)  (L_a(\lambda) + L_b(\mu)) r_{ab}(\lambda-\mu)\Big) =0
\ee
and this concludes our proof. $\Box$

The construction of the Lax pair associated to each one of the integrals of motion utilizing the underlying
classical algebra expressed through (\ref{linear}) is illustrated in what follows. Recall that all the charges in involution are obtained via the transfer matrices
defined in (\ref{tnn}); also the classical equations of motion are expressed as:
\be
{d L(\mu) \over d t} =\Big \{t^{(n)}(\lambda),\ L(\mu) \Big \}. \label{em1}
\ee
To identify the corresponding expression for ${\mathbb A}$ we need to formulate the following expression:
\be
&& \Big \{tr_a\ L_a^n(\lambda),\ L_b(\mu) \Big \} = n\ tr_a\  \Big ( L_a^{n-1}(\lambda)\ \Big \{L_a(\lambda),\ L_b(\mu) \Big \} \Big )
\non\\ && = n \ tr_a \Big (L_a^{n-1}(\lambda) \Big [r_{ab}(\lambda-\mu),\ L_a(\lambda) + L_b(\mu) \Big] \Big ) \non\\
&& n\ tr_a \Big (L_a^{n-1}(\lambda)\ r_{ab}(\lambda- \mu) \Big )\ L_b(\mu) - L_b(\mu)\ n\ tr_a \Big (L_a^{n-1}(\lambda)\
r_{ab}(\lambda- \mu) \Big ). \label{final}
\ee
Comparing (\ref{zero}) with (\ref{em1}) through (\ref{final}) we conclude that \cite{sts, BBT}
\be
{\mathbb A}(\lambda,\ \mu) = n\ tr_a \Big (L_a^{n-1}(\lambda)\ r_{ab}(\lambda- \mu) \Big ).  \label{aa}
\ee
In the case where the $r$-matrix is the Yangian matrix
\cite{yang}
\be
r(\lambda) ={1\over \lambda}{\cal P} \label{yangian}
\ee
${\cal P}$ the permutation operator $~{\cal P}\ (\vec a \otimes \vec b) = \vec b \otimes \vec a$,  the expression
(\ref{aa}) is modified as:
\be
{\mathbb A}(\lambda) = {n \over \lambda -\mu}L^{n-1}(\lambda).
\ee
\\
\textbf{\underline{Example.}} Let us discuss a typical example associated to linear Poisson structures, that is the classical Toda chain. Consider the classical Toda chain \cite{toda}. The associated  $L$-matrix is given as:
\be
L(\lambda) = u e^{{q_1 - q_N \over 2}} e_{1N} + u^{-1} e^{{q_1-q_N\over 2}}e_{N1} + \sum_{j=1}^N p_j
e_{jj}+\sum_{j=1}^{N-1} e^{{q_{j+1} - q_{j} \over 2}} e_{j j+1} + \sum_{j=1}^{N-1} e^{{q_{j +1}- q_{j} \over 2}} e_{j+1 j} \non\\
\label{ltoda}
\ee
where we define: $u = e^{2\lambda}$, and $(e_{ij})_{kl} = \delta_{ik} \delta_{jl}$.
The Lax operator (\ref{ltoda}) satisfies (\ref{linear}) with the $r$-matrix given by \cite{jimbo, jimbo1}
\be
r(\lambda) = {u_1 +u_2 \over u_1 - u_2} \sum_{j=1}^N e_{jj} \otimes e_{jj} + {1\over u_1 -u_2} \Big (u_1\sum_{k > j}^N +
u_2 \sum_{k<j}^N\Big ) e_{jk} \otimes e_{kj}
\ee
where $u_i = e^{2 \lambda_i}$.

The momentum and Hamiltonian of the system may be now extracted. This may be achieved by just considering the first two
generating functions $tr L$, $tr L^2$ and expand in powers of $u$. It is convenient for our purposes here to express the
$L$-matrix in a more compact form as:
\be
L(\lambda) \propto u^2 A + u B +D
\ee
then
\be
tr L(\lambda) \propto u^2\ tr A + u\ tr B + tr D
\ee
but due to the structure of the matrices $A,\ B,\ D$ (\ref{ltoda}) we immediately conclude:
\be
I_{1} = tr B= \sum_{j=1}^N p_j.
\ee
To obtain the second integral of motion we expand
\be
trL^2(\lambda) \propto u^4\ tr (A^2) + u^3\ tr (AB + BA) + u^2\ tr (AD+ DA+ B^2) + u\ tr (DB +BD)+  tr (D^2). \non\\
\ee
Recalling the form of the involved matrices via (\ref{ltoda}) we conclude that the first non-trivial contribution,
which essentially corresponds to the Hamiltonian, is given by the zero order terms, i.e.
\be
I_{2} = - {1\over 2} tr (AD +DA + B^2) = \ldots = -{1\over 2} \sum_{j=1}^N p_j^2 - \sum_{j=1}^N e^{q_{j+1} - q_j}.
\ee
It is clear from the preceding discussion that by construction
\be
\Big \{I_1,\ I_2\Big \} =0.
\ee
We shall come back to the Toda model in a subsequent section providing a ``dual'' description of the model due to Sklyanin \cite{sklyaninlect}-\cite{sklyaninlect2}.
\\
\\
\textbf{\underline{Exercise 3.1.}} Derive the exchange relations among $q_j,\ p_j$.\\
\textit{Tip}: use the following useful identity:
\be
e_{ij}\ e_{kl} = \delta_{jk}\ e_{il}.
\ee

\section{Quadratic Poisson structure: the discrete case}
Our main aim in this and the next section is to present a systematic means for constructing discrete and continuum integrable theories \cite{ftbook, ftbook1}.
Based on solely algebraic considerations we shall provide the general setting for building such systems as well as explain
in detail how the associated charges in involution together with the corresponding Lax pairs may be extracted from the generating
functional i.e. the transfer matrix of the system. We shall first focus on one-dimensional discrete integrable models and then shall
go on and deal with 1+1 dimensional integrable field theories. Both discrete and continuum integrable theories share the same
underlying algebra described by certain quadratic relations, which will be introduced below.

\subsection{Local integrals of motion}
Recall first the auxiliary linear problem in the case of discrete integrable models based on quadratic Poisson algebras. Lax representation of classical dynamical
evolution equations \cite{lax} is one key ingredient in the theory of classical
integrable systems \cite{ftbook}--\cite{ZSh} together with the associated notion of classical $r$-matrix \cite{skl, sts}.
Introduce the Lax pair ($L,\ {\mathbb A}$) for discrete integrable models \cite{abla}, and the associated auxiliary problem (see e.g.
\cite{ftbook})
\be
&& \psi_{n+1} = L_n\ \psi_n \non\\
&& \dot{\psi}_n = {\mathbb A}_n\ \psi_n.
\label{lat}
\ee
From the latter equations one may immediately obtain the discrete zero curvature condition:
\be
\dot{L}_n = {\mathbb A}_{n+1}\ L_n - L_n\ {\mathbb A}_n.
\label{zero2}
\ee
The monodromy matrix arises from the first equation (\ref{lat}) (see e.g. \cite{ftbook, ftbook1})
\be
T_{a}(\lambda) = L_{aN}(\lambda)\ L_{a N-1}(\lambda) \ldots L_{a1}(\lambda) \label{trans0}
\ee
where the index $a$ denotes the auxiliary space, and the
indices $1, \ldots , N$ denote the sites of the one dimensional classical discrete model. Expansion of the transfer matrix in powers of
the spectral parameter gives rise to the charges in involution.
Usually as will be transparent in the examples that follow the $\ln$ of the transfer matrix provides the {\it local} integrals of motion.

Consider a skew symmetric classical $r$-matrix, which is a solution of the classical Yang-Baxter equation, and let $L$ satisfy
the associated Sklyanin bracket
\be
\Big \{ L_{an}(\lambda),\ L_{bm}(\mu) \Big \} = \Big [
r_{ab}(\lambda-\mu),\ L_{an}(\lambda)\ L_{bm}(\mu) \Big ] \delta_{nm}. \label{clalg}
\ee
The following proposition for the monodromy matrix may be then stated.
\\
\\
\textbf{\underline{Proposition 4.1.}} Show that the monodromy matrix defined in (\ref{trans0}) also satisfies the quadratic
algebra (\ref{clalg}):
\be
\Big \{T_a(\lambda),\ T_b(\lambda) \Big \} = \Big [ r_{ab}(\lambda -\mu),\ T_a(\lambda)\ T_b(\mu) \Big ]. \label{falg}
\ee
\\
\textit{\underline{Proof}}: From the LHS of the latter equation and the definition of the monodromy matrix:
\be
&& \Big \{L_{aN}(\lambda) \ldots L_{a1}(\lambda),\ L_{bN}(\mu) \ldots L_{b1}(\mu) \Big \} =  \non\\ && \sum_n
L_{aN}(\lambda) \ldots L_{an+1} L_{bN}(\mu) \ldots L_{bn+1}(\mu) \Big \{L_{an}(\lambda), \ L_{bn}(\mu) \Big \}
L_{an-1}(\lambda) \ldots L_{a1}(\lambda) L_{b n-1}(\mu) \ldots L_{b1}(\mu) = \non\\&& \sum_n L_{aN}(\lambda)
\ldots L_{an+1} L_{bN}(\mu) \ldots L_{bn+1}(\mu) \Big [r_{ab},\ L_{an}(\lambda) L_{bn}(\mu) \Big ]L_{an-1}(\lambda) \ldots
L_{a1}(\lambda) L_{b n-1}(\mu) \ldots L_{b1}(\mu) = \non\\
&&\sum_{n}L_{aN}(\lambda) \ldots L_{an+1} L_{bN}(\mu) \ldots L_{bn+1}(\mu)r_{ab} L_{an}(\lambda) \ldots L_{a1}(\lambda)
L_{b n}(\mu) \ldots L_{b1}(\mu) -\non\\
&& \sum_{n}L_{aN}(\lambda) \ldots L_{an} L_{bN}(\mu) \ldots L_{bn}(\mu)r_{ab} L_{an-1}(\lambda) \ldots L_{a1}(\lambda)
L_{b n-1}(\mu) \ldots L_{b1}(\mu),
\ee
but notice two consecutive terms with opposite signs cancel each other, so the only terms that survive in the sum are
the very first and the last ones. Thus the expression above becomes:
\be
\Big \{ T_a(\lambda),\ T_b(\lambda) \Big \} = r_{ab}(\lambda-\mu)\ T_a(\lambda)\ T_b(\mu) - T_a(\lambda)\ T_b(\mu)\ r_{ab}(\lambda -\mu),
\ee
which concludes our proof. $\Box$
\\
\\
\textbf{\underline{Proposition 4.2.}} Define the transfer matrix of the discrete system as:
\be
t(\lambda) =  tr_a T_a(\lambda).
\ee
Then show that it provides the charges in involution, that is:
\be
\Big \{ t(\lambda),\ t(\lambda') \Big \}=0. \label{transfer}
\ee
\\
\textit{\underline{Proof}}:
The proof is straightforward based on the fundamental quadratic algebra defined by (\ref{falg}), and the definition of the transfer matrix:
\be
\Big \{ t(\lambda),\ t(\lambda') \Big \} &=& \Big \{tr_a T_a(\lambda),\ tr_bT_b(\lambda') \Big \} = tr_{ab} \Big \{T_a(\lambda),\
T_b(\lambda') \Big \} \non\\
&=& tr_{ab} \Big [r_{ab}(\lambda - \lambda'),\ T_a(\lambda)\ T_b(\lambda') \Big ]
\non\\ &=& tr_{ab} \Big (r_{ab}(\lambda -\lambda')\ T_a(\lambda)\ T_b(\lambda')- T_a(\lambda)\ T_b(\lambda')\ r_{ab}(\lambda -\lambda')
\Big ) =0
\ee
and this concludes our proof. $~\Box$

\subsection{Lax pair construction}
Let us now briefly review how the Lax pair associated to each local integral of motion is derived via the
$r$-matrix formulation (see also \cite{ftbook, BBT}). Recall the Lax pair ($L,\ {\mathbb A}$) for discrete integrable models, and the
associated discrete auxiliary linear problem (\ref{lat}).

First some necessary notation is introduced. We define for $i > j$:
\be
T_a(i,j;\lambda)= L_{ai}(\lambda)\ L_{a i-1}(\lambda) \ldots L_{aj}(\lambda).
\ee
\\
\textbf{\underline{Proposition 4.3.}} The operator ${\mathbb A}(\lambda)$ of the Lax pair $L,\ {\mathbb A}$ is identified as:
\be
{\mathbb A}_n(\lambda, \mu) = t^{-1}(\lambda)\
tr_{a}\ \Big  [ T_a(N,n;\lambda)\ r_{ab}(\lambda -\mu)\ T_a(n-1,1;\lambda) \Big ]. \label{laf}
\ee
\\
\textit{\underline{Proof:}}
To be able to construct the Lax pair  we should first  formulate the following Poisson structure \cite{ftbook}:
\be
&&\Big \{ T_a(\lambda),\ L_{bj}(\mu) \Big \} = L_{aN}(\lambda)  \ldots L_{an+1}(\lambda) \Big \{ L_{an}(\lambda),\
L_{bn}(\mu) \Big \}L_{an-1}(\lambda) \ldots  L_{a1}(\lambda)
\non\\ &&= L_{aN}(\lambda)  \ldots L_{an+1}(\lambda) \Big ( r_{ab}(\lambda- \mu) L_{an}(\lambda)L_{bn}(\mu) -
L_{an}(\lambda)L_{bn}(\mu)r_{ab}(\lambda -\mu) \Big )L_{an-1}(\lambda) \ldots L_{a1}(\lambda) \non\\
&&= T_a(N, n+1; \lambda)\ r_{ab}(\lambda - \mu)\ T_a(n,1;\lambda)\ L_{bn}(\mu) - L_{bn}(\mu)\ T(N,n;\lambda)\ r_{ab}(\lambda -\mu)\ T_a(n-1,1;\lambda). \non\\
\ee
It then immediately follows for the generating function of the {\it local} integrals of motion:
\be
\Big \{\ln t(\lambda),\ L_{bn}(\mu)\Big \} &=& t^{-1}(\lambda)\ tr_a\Big (T_a(N,n+1;\lambda)\ r_{ab}(\lambda-\mu)\ T_a(n,1;\lambda)
\Big )\ L_{bn}(\mu) \non\\ &-&  L_{bn}(\mu)\ t^{-1}(\lambda)\ tr_a \Big (T_a(N,n;\lambda)\ r_{ab}(\lambda-\mu)\ T_a(n-1,1;\lambda) \Big). \label{local}
\ee
Recalling the classical equation of motion
\be
\dot{L}_n(\mu) = \Big \{\ln t(\lambda),\ L_n(\mu) \Big \}
\ee
and comparing with expression (\ref{local}) we obtain expression (\ref{laf}),
which concludes our proof. $~\Box$

In the special case where the $r$-matrix is $r(\lambda) = {1\over \lambda}{\cal P}$ \cite{yang},
the operator ${\mathbb A}$ of the Lax pair becomes:
\be
{\mathbb A}_n(\lambda, \mu) ={t^{-1}(\lambda) \over \lambda -\mu}\ T(n-1, 1;\lambda)\ T(N,n;\lambda). \label{ayangian}
\ee
Expansion of the latter expression in powers of ${1 \over \lambda}$ (or $\lambda$) gives rise to the time components of the Lax pairs corresponding to each local integral of motion:
\be
{\mathbb A}(\lambda,\ \mu) = \sum_{j} {{\mathbb A}^{(j)}(\mu) \over \lambda^j}.
\ee

\section{Examples of discrete integrable models}

\subsection{The discrete NLS system or DST model}

The construction presented in the previous section will be exemplified with the study of
the discrete version of the NLS model. This model is also know as
discrete-self-trapping (DST) model describing the non-linear behavior of small of big molecules
(see \cite{nls1}--\cite{sklyaninlect2}).
We introduce the associated Lax operator given by (see e.g. \cite{kundura, kundura1}):
\be
L_{aj}(\lambda) &=& \lambda D_j + A_j \non\\&=&  \begin{pmatrix}
   \lambda +{\mathbb N}_j  & {\mathrm x}_j \\
    - {\mathrm X}_j & 1
\end{pmatrix} \ \label{lmatrix}
\ee where ${\mathbb N}_j = 1-  {\mathrm x}_j  {\mathrm X}_j$. It also turns out, via the quadratic algebraic relation satisfied by the $L$ operator, that $ {\mathrm x},\  {\mathrm X}$ are canonical variables, i.e.
\be
\{ {\mathrm x}_i,\  {\mathrm X}_j\} = \delta_{ij}.
\ee

To derive the {\it local} integrals of motion one should expand the $\ln t(\lambda)$ in powers of ${1\over \lambda}$.
Let us expand the monodromy matrix:
\be
T(\lambda \to \infty) & \propto & D_N\ldots D_1  + {1\over \lambda} \sum_{i=1}^N D_N \dots D_{i+1} A_i D_{i-1}
\ldots D_1
\non\\&+& {1\over \lambda^2} \sum_{i>j} D_N \ldots D_{i+1} A_i D_{i-1} \ldots D_{j+1} A_j \ldots D_1
\non\\
&+& {1\over \lambda^3} \sum_{i>j>k} D_N \ldots D_{i+1} A_i D_{i-1} \ldots D_{j+1} A_j
\ldots D_{k+1} A_k \ldots D_1 \non\\ &+& \ldots
\ee
Taking into account the latter expansion we conclude:
\be
\ln t(\lambda \to \infty) \propto {1 \over \lambda} I_1 + {1\over \lambda^2} I_2 + {1\over \lambda^3} I_3 + \ldots
\ee
where the extracted integrals of motion have the following familiar form (see also e.g. \cite{kundura, kundura1, doikouit})
\be
&& I_1 = \sum_{i=1}^N {\mathbb N}_i, \non\\
&& I_2 = - \sum_{i=1}^N  {\mathrm x}_{i+1}  {\mathrm X}_i - {1\over 2} \sum_{i=1}^N {\mathbb N}_i^2 \non\\
&& I_3 = -\sum_{i=1}^N  {\mathrm x}_{i+2}{\mathrm X}_i + \sum_{i=1}^N ({\mathbb N}_i+{\mathbb N}_{i+1}) {\mathrm x}_{i+1} {\mathrm X}_i
+{1\over 3}\sum_{i=1}^N{\mathbb N}_i^3.
\ee
The latter provide the first three integrals of motion (number of particles, momentum and Hamiltonian respectively) of the whole hierarchy for the NLS model. The continuum limits of the above quantities provide the corresponding integrals of motion of the continuum NLS model \cite{kundura, kundura1, doikouit} as will be also transparent later in the text. It is worth noting that the latter expressions are valid in the quantum case as well (see e.g. \cite{kundura, kundura1, doikouit}).

The associated Lax pairs may be now identified.
As already noted expansion of the expression (\ref{ayangian}) in powers of ${1\over \lambda}$ provides the Lax pairs associated to each one of the
local integrals of motion (see also \cite{avandoikou}):
\be
&&  {\mathbb A}_j^{(1)}(\mu) =  \begin{pmatrix}
   1 & 0 \\
    0 & 0
\end{pmatrix}, ~~~~
 {\mathbb A}_j^{(2)}(\mu) =\begin{pmatrix}
   \mu &  {\mathrm x}_j \\
    - {\mathrm X}_{j-1} & 0
\end{pmatrix}, \non\\
&&  {\mathbb A}_j^{(3)}= \begin{pmatrix}
   \mu^2  + {\mathrm x}_j  {\mathrm X}_{j-1} & \mu  {\mathrm x}_j - {\mathrm x}_j {\mathbb N}_j +  {\mathrm x}_{j+1}\\
    -\mu  {\mathrm X}_{j-1} +  {\mathrm X}_{j-1}{\mathbb N}_{j-1}- {\mathrm X}_{j-2} & - {\mathrm x}_j {\mathrm X}_{j-1}
\end{pmatrix}. \label{bulka}
\ee

Both the Lax pair via the zero curvature condition, and the Hamiltonian description give rise to the same equations of motion.
Consider for instance
the equations of motion associated to $I_3$ (and the Lax pair $L,\  {\mathbb A}^{(3)}$). Indeed from
\be
\dot{ {\mathrm x}}_j = \{I_3,\  {\mathrm x}_j\}, ~~~~\dot{ {\mathrm X}}_j= \{I_3,\  {\mathrm X}_j \},
\ee
and via the zero curvature condition for the pair $L,\ {\mathbb A}^{(3)}$ we obtain the following set of difference equations:
\be
\dot{ {\mathrm x}}_j &=&  {\mathrm x}_{j+2} - 2 {\mathrm x}_{j+1} {\mathbb N}_j -   {\mathrm x}_{j+1} {\mathbb N}_{j+1 }+  {\mathrm x}_j {\mathbb N}_j^2 +  {\mathrm x}_j^2  {\mathrm X}_{j-1} +  {\mathrm x}_{j+1}\non\\
\dot{ {\mathrm X}}_j &=& - {\mathrm X}_{j-2} +2 {\mathrm X}_{j-1} {\mathbb N}_j +  {\mathrm X}_{j-1} {\mathbb N}_{j-1} - {\mathrm X}_j {\mathbb N}_j^2 - {\mathrm X}_j^2  {\mathrm x}_{j+1}-  {\mathrm X}_{j-1}.
\label{bulke}
\ee
With this we conclude our brief review on the periodic discrete NLS model.

\subsection{The Toda model}
The Toda model may be also obtained as a suitable limit of the DST model. Then the harmonic oscillator algebra formed by $( {\mathrm x},\  {\mathrm X},\  {\mathrm x} {\mathrm X})$ reduces to the Euclidean algebra generated by $(e^q,\ p)$ (see \cite{sklyaninlect}--\cite{sklyaninlect2}). The associated Lax operator for the Toda model is given by:
\be
L_{aj}(\lambda) =  \begin{pmatrix}
   \lambda -p_j  & e^{q_j} \\
    -e^{-q_j} & 0
\end{pmatrix} \ \label{tomatrix}.
\ee
\\
\textbf{\underline{Exercise 5.1.}} Show that the first two local integrals of motion (momentum and Hamiltonian) for the Toda chain are given by the following expressions:
\be
I_1 &=& \sum_{j=1}^N p_j, \non\\
I_2 &=& -{1\over 2}\sum_{j=1}^N p_j^2 - \sum_{j=1}^N e^{q_{j+1} - q_j}.
\ee
Extract also the associated equations of motion.

Recall the linear description of the Toda chain presented in section 3; the expressions for the momentum and Hamiltonian clearly coincide.
\\
\\
\textbf{\underline{Exercise 5.2.}} Extract the Lax pair associated to the Hamiltonian $I_2$ of the Toda model, and show that is given as
\be
{\mathbb A}_j(\lambda) = \begin{pmatrix}
   \lambda   & e^{q_j} \\
    -e^{q_{j-1}} & 0
\end{pmatrix} \,.
\ee
Show also that the corresponding equations of motion are given by:
\be
p_j  = \dot{q}_j, ~~~~~\ddot{q}_j = e^{q_{j+1}-q_j} -e^{q_{j}-q_{j-1}}.
\ee

\section{Quadratic Poisson structure: the continuous case}

\subsection{Local integrals of motion}
The basic notions regarding the Lax pair and the zero curvature condition for a continuous integrable model are reviewed
following essentially \cite{ftbook}. Define $\Psi$ as being a solution of the following set of equations (see e.g. \cite{ftbook})
\be
&&{\partial \Psi \over \partial x} = {\mathbb U}(x,t, \lambda)  \Psi \label{dif1}\\
&& {\partial \Psi \over \partial t } = {\mathbb
V}(x,t,\lambda) \Psi \label{dif2}
\ee
${\mathbb U},\ {\mathbb V}$ being in general $n \times n$ matrices with entries defined as
functions of complex valued dynamical fields, their derivatives, and the spectral parameter $\lambda$.
The monodromy matrix from
(\ref{dif1}) may be written as:
\be
T(x,y,\lambda) = P \exp \Big \{ \int_{y}^x {\mathbb U}(x',t,\lambda)dx' \Big \}.
\label{trans}
\ee
The fact that $T$ also satisfies equation (\ref{dif1}) will be extensively used to get the relevant integrals of
motion. Compatibility conditions of the two differential equations (\ref{dif1}), (\ref{dif2}) lead to the zero curvature condition
\cite{AKNS}--\cite{ZSh}
\be
\dot{{\mathbb U}} - {\mathbb V}' + \Big [{\mathbb U},\ {\mathbb V} \Big ]=0, \label{zecu}
\ee
giving
rise to the corresponding classical equations of motion of the system under consideration.

Hamiltonian formulation of the equations of motion is available under the $r$-matrix approach. In this picture the underlying
classical algebra is manifestly analogous to the quantum case. The existence of the Poisson structure for ${\mathbb U}$ realized by
the classical $r$-matrix, satisfying the classical Yang-Baxter equation, guarantees the integrability of the classical
system. Indeed, assuming that the operator ${\mathbb U}$ satisfies the following ultra-local form of Poisson brackets
\be
\Big \{{\mathbb U}_a(x, \lambda),\ {\mathbb U}_b(y, \mu) \Big \} = \Big [r_{ab}(\lambda - \mu),\ {\mathbb U}_a(x, \lambda) +{\mathbb
U}_b (y,\mu) \Big ]\ \delta(x-y), \label{ff0}
\ee
then $T(x,y,\lambda)$ satisfies (\ref{clalg}),
and consequently one may readily
show for a system on the full line:
\be
\Big \{\ln tr\{T(x,y,\lambda_1)\},\ \ln tr\{T(x,y, \lambda_2)\} \Big\}=0
\ee
i.e. the system is integrable, and the charges in involution --local integrals of motion-- are obtained by expansion of the generating
function $\ln tr\{T(x,y,\lambda)\}$, based essentially on the fact that $T$ satisfies (\ref{dif1}).

\subsection{Lax pair construction}
The construction of the time component ${\mathbb V}$ of the Lax pair associated to the local integrals of motion is outlined below. The first step
is to formulate the following expression:
\be
\Big \{ T_a(L, -L,\lambda),\ {\mathbb U}_b(x, \mu) \Big \} = {\partial {\mathbb M}_a(x, \lambda, \mu) \over \partial x} + \Big
[{\mathbb M}_a(x, \lambda, \mu),\ {\mathbb U}_b(x,\ \mu ) \Big] \label{last}
\ee
where we define
\be
{\mathbb M}_a(x,\lambda, \mu)
&=& T_a(L, x, \lambda)\ r_{ab}(\lambda -\mu)\ T_a(x, -L, \lambda). \label{mm}
\ee
For more details on the proof we refer the interested reader to \cite{ftbook}, but the latter is an immediate consequence of the process followed in the discrete case after considering the continuum limit. We shall discuss in more detail the continuum limit of integrable discrete models in a subsequent section.

Bearing also in mind the
definition of $t(\lambda)$, and (\ref{last}) it is straightforward to show:
\be
 \Big \{ \ln\ t(\lambda),\ {\mathbb U}(x, \mu) \Big
\} = {\partial {\mathbb V}(x,\lambda, \mu) \over \partial x} + \Big [ {\mathbb V}(x,\lambda, \mu),\ {\mathbb U}(x, \mu) \Big ],
\label{defin}
\ee
where we define
\be {\mathbb V}(x,\lambda, \mu) = t^{-1}(\lambda) \ tr_a  {\mathbb M}_a(x , \lambda,
\mu). \label{final1}
\ee

Special examples of classical integrable field theories are presented in the subsequent section.

\section{Examples of integrable field theories}
\subsection{The NLS model}
A particular example associated to the rational $r$-matrix \cite{yang}, that is the non-linear Schrodinger (NLS) model will be investigated below. The Lax operator for the system is given by
the following expressions \cite{ftbook, foku}:
 \be
{\mathbb U} = \begin{pmatrix}
   {\lambda \over 2}  & \bar \psi \\
    \psi & - {\lambda \over 2}
\end{pmatrix} \ ,  \label{lax0}
\ee
and it is shown via the linear algebra (\ref{ff0}) that $\psi,\ \bar \psi$
satisfy
\be \Big \{\psi(x),\ \bar \psi(y)
\Big \}=  \delta(x-y).
\ee

Our main aim is to extract the first couple of local integrals of motion via the expansion of $\ln t(\lambda)$ in powers of ${1\over \lambda}$. We shall start this construction recalling that the classical monodromy matrix satisfies the fundamental quadratic algebra (\ref{falg}).
Consider also the following ansatz for the monodromy matrix:
\be
T(x,y,\l) = (1+W(x,\l))\ e^{Z(x,y,\l)}\ (1+W(y,\l))^{-1},
\label{transatz}
\ee
$W$ and $Z$ are purely off-diagonal and diagonal matrices respectively. We also assume that $W,\ Z$
are expressed as:
\be
W(x,\l)=\sum_{n=1}^{\infty}{ W_n(x) \over \lambda^n}, ~~~~~ Z(x,y,\l)=\sum_{n=-1}^{\infty}{Z_n(x,y)\over  \lambda^n}.
\label{WZexp}
\ee
Our aim henceforth is to identify the elements $W_n,\ Z_n$, and hence the integrals of motion.
It is technically convenient to split the Lax operator into a diagonal and an off-diagonal part as
\be
\mathbb{U} = \mathbb{U}_d + \mathbb{U}_a \equiv \frac{\lambda}{2}
\begin{pmatrix}
  1 & 0 \cr
  0 & -1
\end{pmatrix}
+
\begin{pmatrix}
  0 & \bar \psi \cr
  \psi & 0
\end{pmatrix}.
\ee
Substituting the ansatz (\ref{transatz}) into \eqn{dif1}, and splitting the
resulting equation into a diagonal and an off-diagonal part one obtains
\be
&& \frac{\partial W}{\partial x} + W \mathbb{U}_d - \mathbb{U}_d W +  W \mathbb{U}_a W- \mathbb{U}_a =0,\cr
&& \frac{\del Z}{\del x} = \mathbb{U}_d + \mathbb{U}_a W. \label{split0}
\ee

We may now solve the latter equations for each order and determine all $W^{(n)},\ Z^{(n)}$ and consequently the local integrals of motion. More precisely, recall the generating function of the local integrals of motion
\be
{\cal G}(\lambda)=  \ln \Big (tr T(\lambda)\Big )
\ee
due to the ansatz (\ref{transatz}) we can substitute the monodromy matrix accordingly and obtain:
\be
{\cal G}(\lambda)= \ln tr \Big [ (1+W(L))\ e^{Z(L, -L)}\ (1+W(-L))^{-1} \Big ],
\ee
but due to the choice of periodic boundary conditions
we conclude that:
\be
{\cal G}(\lambda) = \ln tr \Big [e^{Z(L, -L, \lambda)} \Big ].
\ee
Let us first evaluate the first couple of $W^{(i)}$'s,
\be
&& W^{(1)} = \begin{pmatrix}
    &       -\bar \psi(x) \\
    \psi(x) &
\end{pmatrix} \ \label{1lmatrix}, ~~~~~W^{(2)} = \begin{pmatrix}
    &       -\bar \psi'(x) \\
    -\psi'(x) &
\end{pmatrix} \ \label{2lmatrix} \non\\
&& W^{(3)} = \begin{pmatrix}
    &       -\bar \psi''(x) + |\psi(x)|^2 \bar \psi(x)\\
    \psi''(x)- |\psi(x)|^2  \psi(x)&
\end{pmatrix}. \label{3lmatrix}
\ee

Similarly, through (\ref{split0}) the diagonal elements $Z^{(i)}$ are given as:
\be
&& Z^{(-1)} = \begin{pmatrix}
  L  &        \\
     & -L
\end{pmatrix} \ \label{4lmatrix}, ~~~~~Z^{(1)} = \begin{pmatrix}
   \int_{-L}^L dx\ \psi(x) \bar \psi(x)  &       \\
  &-\int_{-L}^L dx\ \psi(x) \bar \psi(x)
\end{pmatrix} \ \label{5lmatrix}\non\\
&& Z^{(2)} = \begin{pmatrix}
  -\int_{-L}^L dx\ \psi'(x) \bar \psi(x)  &     \\
     & -\int_{-L}^L dx\ \psi(x) \bar \psi'(x)
\end{pmatrix} \ \label{6lmatrix} \non\\
&& Z^{(3)} = \begin{pmatrix}
  \int_{-L}^L dx\ \Big ( \psi''(x) \bar \psi(x) - |\psi(x)|^4  \Big )  &       \\
            & -\int_{-L}^L dx\ \Big (\bar \psi''(x) \psi(x)- |\psi(x)|^4  \Big )
\end{pmatrix}. \label{7lmatrix}
\ee

The expansion of the generating function ${\cal G}$ in powers of ${1\over \lambda}$ provides the local integrals of motion of the model under consideration. Note that due to the fact that for $\lambda \to \infty$
the leading contribution comes from $Z^{(-1)}_{11}$, all the local integrals of motion are given essentially by the quantities $Z_{11}^{(n)}$. More precisely the first three integrals of motion --number of particles, momentum and Hamiltonian -- may be respectively expressed as
\be
{\cal N} &=& \int_{-L}^L dx\ \psi(x) \bar \psi(x) \non\\
{\cal P} &=& {1\over 2} \int_{-L}^L dx\ \Big ( \bar \psi'(x) \psi(x) - \bar \psi(x) \psi'(x) \Big ) \non\\
{\cal H} &=& -\int_{-L}^L dx\ \Big ( |\psi(x)|^4  +\psi'(x) \bar
\psi'(x)\Big ). \label{nlh}
\ee

The next natural step is the construction of the relevant Lax pair, in particular the associated ${\mathbb V}$ operators.
Recall the generic expression of the ${\mathbb V}$-operator derived in the previous section:
\be
{\mathbb V}(x, \lambda, \mu)= t^{-1}(\lambda)\  tr_a \Big ( T_a(L, x, \lambda)\  r_{ab}(\lambda -\mu)\ T_a(x, -L, \lambda)  \Big ). \label{sp}
\ee
Recall that in this case the relevant $r$-matrix is the Yangian, so substituting $r(\lambda) = {{\cal P} \over \lambda}$ in the latter expression we end up:
\be
{\mathbb V}(x, \lambda, \mu) = {t^{-1}(\lambda) \over \lambda -\mu}\ T(x, -L,\lambda)\ T(L,x, \lambda). \label{vv2}
\ee
Ultimately we wish to expand $T(\lambda)$ in powers of $\lambda^{-1}$ in order to
determine the relevant Lax pairs from (\ref{vv2}).
Recall the ansatz for the classical monodromy matrix ({\ref{transatz}), after substituting (\ref{transatz}) in the expression (\ref{vv2}) and recalling that the leading contribution comes from the $Z_{11}$ term, we conclude:
\be
{\mathbb V}(\lambda)= {1 \over \lambda -\mu} (1 + W(x, \lambda))\ {\cal D}\ (1+W(x, \lambda))^{-1} \label{vfinal}
\ee
where we define
\be
{\cal D} =  \begin{pmatrix}
   1  & 0     \\
   0  & 0
\end{pmatrix}.
\ee
We then expand expression (\ref{vfinal}) in powers of ${1\over \lambda}$, and thus we identify the corresponding ${\mathbb V}$-operators.
Indeed,
\be
{\mathbb V}(\lambda, \mu) =  \sum_{n=1}^{\infty}{{\mathbb V}^{(n)} \over \lambda^{n}}.
\ee
We provide here the quantities associated to the first three integrals of motion:
\be
{\mathbb V}^{(1)} &=& {\cal D},  ~~~~~~{\mathbb V}^{(2)}(\mu) = \begin{pmatrix}
   \mu  & \bar \psi    \\
   \psi  & 0
\end{pmatrix} \non\\
{\mathbb V}^{(3)}(\mu)& =& \begin{pmatrix}
   \mu^2 - \psi\ \bar \psi & \mu \bar \psi  + \bar \psi'   \\
   \mu \psi -\psi' & \psi\ \bar \psi
\end{pmatrix}, ~~~\ldots
\ee
Details on the computation of these quantities are left as an exercise to the interested reader.

It is now straightforward to obtain the equations of motion relevant to each integral of motions. Focus for instance on the Hamiltonian of the system ${\cal H}$ (\ref{nlh}). From the zero curvature condition for the pair ${\mathbb U},\ {\mathbb V}^{(3)}$ one has:
\be
\dot {\mathbb U} - {\mathbb V}^{(3)'} + \Big [{\mathbb U},\  {\mathbb V}^{(3)} \Big ] =0,
\ee
which leads to the following familiar equations of motion
\be
\dot{\psi}(x) &=& {\partial^2 \psi(x) \over \partial x^2} - 2 |\psi(x)|^2 \psi(x)\non\\
\dot{\bar \psi}(x) &=& {\partial^2 \bar \psi(x) \over \partial x^2} - 2 |\psi(x)|^2 \bar \psi(x).
\ee
It is clear that the same equations of motion are entailed
\be
\dot{\psi}(x) = \Big \{ {\cal H},\ \psi(x) \Big \}, ~~~~\dot {\bar \psi}(x) = \Big \{ {\cal H},\ \bar \psi(x) \Big \},
\ee
and this of course verifies the consistency of the whole process. Results on the generalized vector $gl(n)$ NLS model are presented in \cite{doikouit}.
\\
\\
{\underline{{\bf Exercise 7.1.}}
Introduce the isotropic Landau-Lifshitz (classical Heisenberg) model with Lax operator (see e.g. \cite{ftbook, doikoukaraiskos}):
\be
\mathbb{U}(x) =  \frac{1}{\l}
\begin{pmatrix}
 \frac{S_3}{2} & S^- \cr
 S^+ & -\frac{S_3}{2}
\end{pmatrix}
\equiv \frac{1}{2\l}\mathcal{S}.
\label{U_bulk}
\ee
satisfying the linear algebra with $r$ matrix being the Yangian, and thus the Poisson structure of the phase space for the physical quantities $S_i(x)$ is given by the
Poisson brackets
\be
\{S_a(x),S_b(y)\} = 2i\varepsilon_{abc}S_c(x)\d(x-y),
\ee
where $\varepsilon_{abc}$ is the totally antisymmetric Levi-Civita tensor with value $\varepsilon_{123} = 1$.
Based on the knowledge of the Lax operator ${\mathbb U}$:\\
{\bf (i)} Determine the first two integrals of motion from the expansion of the $\ln tr T(\lambda)$ in powers of $\lambda$:
\be
\ln tr T(\lambda) = \sum_{n=0}^{\infty} {I^{(n)} \over \lambda^n}.
\ee
These are the momentum and Hamiltonian and are given as:
\be
I^{(0)} &\propto& {\cal P} =\int_{-L}^L \frac{S_1\frac{\del S_2}{\del x}-S_2 \frac{\del S_1}{\del x}}{1+S_3}dx, \non\\
I^{(1)} &\propto & {\cal H} =  - \frac{1}{4}\int_{-L}^L \left(\left(\frac{\del S_1}{\del x}\right)^2 + \left(\frac{\del S_2}{\del x}\right)^2
+ \left(\frac{\del S_3}{\del x}\right)^2 \right)dx.
\ee
{\bf (ii)} Derive the ${\mathbb V}$-operator associated to the Hamiltonian and show that is of the form:
\be
\mathbb{V}(x) = \frac{1}{2  \l^2} \mathcal{S} - \frac{1}{2\l} \frac{\del\mathcal{S}}{\del x}
\mathcal{S}.
\label{V_bulk}
\ee
{\bf (iii)} Show that the equations of motion relevant to the Hamiltonian are give by:
\be
\frac{\del \vec{S}}{\del t}=i \vec{S} \wedge {\partial^2 \vec{S} \over \partial x^2}.
\label{e.o.m.}
\ee

\subsection{Affine Toda field theories}
We focus on a physically interesting class of integrable models that is the affine Toda field theories.
The first model in this hierarchy is the well known sine-Gordon model. Suitable massless limits of these theories give rise to the Liouville model and to its higher rank generalizations (massless ATFTs). We shall exemplify our study by considering the first two models of the hierarchy, i.e. the $A_1^{(1)}$ (sine(sinh)-Gordon) and the $A_2^{(1)}$ ATFT's, as well as the Liouville theory.

\subsubsection{The Kac-Moody ${\hat A}_n^{(1)}$ algebra}
Before we discuss the affine Toda field theories in detail it will be useful for what is described in the subsequent sections to recall the basic definitions regarding the Kac-Moody ${\hat A}_n^{(1)}$ algebra. Let
\be a_{ij} = 2 \delta_{ij} - (\delta_{i\ j+1}+ \delta_{i\ j-1} +\delta_{i1}\ \delta_{jn+1}+\delta_{in+1}\ \delta_{j1}), ~~i,\ j\in \{1, \ldots , n +1\}
\ee
be the Cartan matrix of the affine Lie algebra
${\widehat{sl}_{n+1}}$\footnote{For the $\widehat{sl}_{2}$ case in particular
\be a_{ij} =2\delta_{ij} -2 (\delta_{i1}\ \delta_{j2} +\delta_{i2}\ \delta_{j1}), ~~i,\ j \in \{ 1, 2\}
\ee}
\cite{kac}.
Also define:
\be
[m]!= \prod_{k=1}^{m}\ k,~~~~~\left [ \begin{array}{c}
m \\
n \\ \end{array} \right  ] = {[m]! \over [n]!\ [m-n]!}, ~~~m>n>0.
\ee\\
\\
{\underline {\bf Definition.}} The affine enveloping
algebra $\widehat{sl}_{n+1}$ has the
Chevalley-Serre generators  $e_{i}$, $f_{i}$,
$h_{i}$, $i\in \{1, \ldots, n+1\}$ obeying the defining relations:
\be
&& \Big [h_{i},\ h_{j} \Big]=0\, \qquad \Big [h_{i},\ e_{j} \Big ]=a_{ij}\ e_{j}\,
\qquad \Big [h_{i},\ f_{j} \Big ]
= -a_{ij}\ f_{j}\ , \non\\
&& \Big [e_{i},\ f_{j}\Big ] = \delta_{ij}\ h_{i},
~~~~i,j \in \{ 1, \ldots,n \}
\label{1} \ee
and the Serre relations
\begin{eqnarray}
&& \sum_{n=0}^{1-a_{ij}} (-1)^{n}
\left [ \begin{array}{c}
  1-a_{ij} \\
   n \\ \end{array} \right  ]_{q}
\chi_{i}^{1-a_{ij}-n}\ \chi_{j}\ \chi_{i}^{n} =0, ~~~\chi_{i} \in \{e_{i},\ f_{i} \}, ~~~ i \neq j. \label{chev}
\end{eqnarray}
{\underline {\it Remark:}} The generators $e_{i}$, $f_{i}$, $h_{i}$ for $i\in \{1, \ldots, n\}$ form the $sl_{n+1}$ algebra. Also, $h_{i}= \epsilon_{i} -\epsilon_{i+1}$, where the elements
$ \epsilon_{i}$ belong to $gl_{n+1}$. Recall that $gl_{n+1}$ is derived by adding to $sl_{n+1}$ the elements $ \epsilon_{i}$ $i\in \{1, \ldots, n+1\}$  so that $\sum_{i=1}^{n}\epsilon_{i}$ belongs to the center.
Furthermore, there exist the elements ${\cal E}_{ij} \in gl_{n+1}$ $i\neq j$, with
${\cal E}_{i\ i+1} =e_{i}$, ${\cal  E}_{i+1\ i} =f_{i}$ $i\in \{1, \ldots n-1\}$ and
\be
{\cal E}_{ij} &=& {\cal E}_{ik}\ {\cal E}_{kj}-{\cal E}_{kj}\ {\cal E}_{ik},~~~j \lessgtr k \lessgtr i, ~~~i,j \in \{1, \ldots ,n+1  \}. \label{ee}
\ee
It is clear that  ${\cal E}_{ij} \in gl_{n+1}$ because they can be written solely in terms of the generators $e_{i}$, $f_{i}$, $~i \in \{1, \ldots, n \}$.  $\square$

We also provide explicit expressions of the simple
roots and the Cartan generators for $A_{n}^{(1)}$ \cite{georgi}, in order to express the Kac-Moody algebra in the Cartan-Weyl basis.
The vectors $\alpha_{i} = (\alpha_{i}^{1} \,, \ldots \,,
\alpha_{i}^{n})$ are the simple roots of the Lie algebra of rank
$n+1$ normalized to unity $\alpha_{i} \cdot \alpha_{i} = 1$, i.e.
\be
\alpha_{i} = \Bigl(0 \,,  \ldots \,, 0 \,, -\sqrt{i-1\over 2i}
\,, \stackrel{\stackrel{i^{th}}{\downarrow}} {\sqrt{i+1\over 2i}}
\,, 0 \,,  \ldots \,, 0 \Bigr), ~~~~i \in \{1, \ldots n \}
\ee
Also define the fundamental weights $\mu_{k} = (\mu_{k}^{1} \,,
\ldots \,, \mu_{k}^{n}) \,, \quad  k = 1 \,, \ldots \,, n $ as
(see, e.g., \cite{georgi}).
\be
\alpha_{j} \cdot \mu_{k} = {1\over 2} \delta_{j,k} \,. \label{important}
\ee
The extended (affine)
root $a_{n+1}$  is provided by the relation \be \sum_{i=1}^{n+1}
a_i =0. \ee
The ${\hat A}_{n}^{(1)}$ algebra in the Cartan-Weyl basis
is expressed as:
\be
&&
\Big [H,\ E_{\pm \alpha_i} \Big]= \pm \alpha_i\ E_{\pm \alpha_i},
\non\\ && \Big [E_{\alpha_i},\ E_{-\alpha_i} \Big ] = {2\over
\alpha_i^2}\ \alpha_i \cdot H.
\ee

We provide below the Cartan-Weyl generators in the
defining representation:
\be
E_{\alpha_{i}} &\mapsto& e_{i\ i+1} \,,
\qquad E_{-\alpha_{i}} \mapsto e_{i+1\ i} \,, \qquad E_{\alpha_{n+1}} \mapsto -
e_{n+1\ 1} \,,
\qquad E_{-\alpha_{n+1}} \mapsto - e_{1\ n+1}\non \\
H_{i} &\mapsto& \sum_{j=1}^{n} \mu_{j}^{i} (e_{j j} -e_{j+1\ j+1}) \,,
\qquad i = 1 \,, \ldots \,, n. \label{cartan/weyl/basis}
\ee
For $A_{2}^{(1)}$ in particular we have:
\be
\alpha_1 = (1,\ 0),
~~~\alpha_2 = (-{1\over 2},\ {\sqrt 3 \over 2}),~~~~\alpha_3 =
(-{1\over 2},\ -{\sqrt 3 \over 2})
\ee
define also the following
$3\times 3$ generators \be E_{1} = E_{-1}^t = e_{12}, ~~~~E_2 =
E^t_{-2} = e_{23}, ~~~~E_3= E^t_{-3} = -e_{31},
\ee
where we recall
the matrices $e_{ij}$: $\ (e_{ij})_{kl} = \delta_{ik}\
\delta_{jl}$.
The diagonal Cartan generators $H_{1,2}$ are
\be H_1
={1\over 2} (e_{11} -e_{22}), ~~~~H_2={1\over 2 \sqrt 3}
(e_{11}+e_{22} -2 e_{33}).
\ee

\subsubsection{The $A_n^{(1)}$ ATFT: preliminaries}
Basic notions relevant to the derivation
of the local integrals of motion and the corresponding Lax pairs in ATFT's are reviewed.
Recall first the Lax pair for
a generic $A_{n}^{(1)}$ theory \cite{olive}--\cite{olive4}:
\be
&& {\mathbb V}(x,t,u) = {\beta \over 2}\
\partial_x \Phi \cdot H + {m \over 4}\ \Big (u\ e^{{\beta \over 2}
\Phi \cdot H}\ E_{+}\ e^{- {\beta \over 2} \Phi \cdot H} -{1\over
u}\ e^{-{\beta \over 2}
\Phi \cdot H}\ E_{-}\ e^{{\beta \over 2} \Phi \cdot H}\Big) \non\\
&& {\mathbb U}(x,t,u) = {\beta \over 2}\ \Pi \cdot H + {m \over
4}\ \Big (u\ e^{{\beta \over 2} \Phi \cdot H}\ E_{+}\ e^{- {\beta
\over 2} \Phi \cdot H} +{1\over u}\ e^{-{\beta \over 2} \Phi \cdot
H}\ E_{-}\ e^{{\beta \over 2} \Phi \cdot H}\Big)  \label{lpair}
\ee
$~u=e^{{2\lambda \over n+1}}$ is
the multiplicative spectral parameter, and $\Phi,\ \Pi$ are $n$-vector fields, with components $\phi_i,\
\pi_i,\ i \in \{1, \ldots , n\}$ which are canonical i.e.
\be
\Big \{\phi_i(x),\ \pi_j(y) \Big \} = \delta_{ij}\ \delta(x-y).
\ee
Also define:
\be
E_+ = \sum_{i=1}^{n+1} E_{\alpha_i},
~~~~~E_{-}=\sum_{i=1}^{n+1} E_{-\alpha_i}
\ee
$\alpha_i$ are the
simple roots, $H$ ($n$-vector) and $E_{\pm \alpha_i}$ are the
algebra generators in the Cartan-Weyl basis corresponding to
simple roots defined in (\ref{cartan/weyl/basis}).

The associated classical $r$-matrix is given by
\cite{jimbo, jimbo1}:
\be r(\lambda) = {\cosh(\lambda) \over \sinh
(\lambda)} \sum_{i=1}^{n+1} e_{ii}\otimes e_{ii} + {1\over \sinh
(\lambda)} \sum_{i \neq j =1}^{n+1} e^{ [ sgn(i-j) -(i-j) {2 \over
n+1}  ] \lambda } e_{ij} \otimes e_{ji}. \label{rc}
\ee

Notice that the Lax pair has the following behavior:
\be
{\mathbb V}^t(x, t, -u^{-1}) = {\mathbb V}(x,t, u), ~~~~~{\mathbb
U}^t(x, t, u^{-1}) = {\mathbb U}(x,t,u)
\ee
where $^t$ denotes usual transposition.

To recover the local
integrals of motion of the system under consideration we shall follow the
quite standard procedure, and expand $\ln\ tr T(u)$ in powers of
$u^{-1}$. To expand the transfer matrix and derive the local integrals of motion we
shall need the expansion of $T(x,y, u)$. In what follows in the present section we basically
introduce the necessary preliminaries for such a derivation, and later in the text we
reproduce the known integrals of motion for the $A_{1}^{(1)},\ A_{2}^{(1)}$ ATFT's on the
full line.

Following the logic described in
\cite{ftbook} for the sine-Gordon model, we aim at expressing the part
associated to $E_+$ in ${\mathbb U},$ independently of the fields, after applying a
suitable gauge transformation. More precisely, consider the
following gauge transformation:
\be
T(x, y, u )= \Omega(x)\ \tilde T(x,y, u)\ \Omega^{-1}(y), ~~~~~~~~
\Omega(x) = e^{{\beta \over 2} \Phi(x)\cdot H}. \label{decomp}
\ee
Then from  equation (\ref{dif1}) we obtain that the gauge transformed
operator ${\mathbb U}$ may be expressed as:
\be
\tilde {\mathbb U}(x,t,u) = \Omega^{-1}(x)\ {\mathbb
U}(x,t,u)\ \Omega(x) - \Omega^{-1}(x)\ {d \Omega(x)\over dx}.
\ee
After
implementing the gauge transformations the operator $\tilde
{\mathbb U}$ take the following simple form:
\be
\tilde {\mathbb U}(x,t, u) = {\beta \over 2} \Theta \cdot H
+{m\over 4} \Big ( u E_+ + {1 \over u} X_- \Big ),
\ee
where we define:
\be
\Theta = \Pi -
\partial_x \Phi, ~~~~~X_- =
e^{- \beta \Phi \cdot H}\ E_-\ e^{\beta  \Phi \cdot H},
\ee
$\Theta$ is a $n$-vector with components $\theta_i$, and it is clear that $\tilde T$ now satisfies:
\be
{d \tilde T(x, y, \lambda) \over dx} = \tilde {\mathbb U}(x, \lambda)\ \tilde T(x, y,\lambda). \label{newt}
\ee

Consider now the following ansatz for $\tilde T$ as
$|u| \to \infty$ \cite{ftbook}
\be
\tilde T(x,y,u) = ({\mathbb I}
+W(x, u))\ \exp[Z(x,y,u)]\ ({\mathbb I} +W(y,u))^{-1}, \label{exp0}
\ee
where $W,\ \hat W$ are off diagonal matrices i.e. $~W =
\sum_{i\neq j} W_{ij} E_{ij}$, and $Z,\ \hat Z$ are purely
diagonal $~Z = \sum_{i=1}^{n+1} Z_{ii}E_{ii}$. Also
\be Z(u) =
\sum_{k=-1}^{\infty} {Z^{(k)} \over u^{k}}, ~~~~W_{ij} =
\sum_{k=0}^{\infty}{W^{(k)} \over u^k}. \label{expa}
\ee
Inserting
the latter expressions (\ref{expa}) in (\ref{newt}) one may
identify the coefficients $W_{ij}^{(k)}$ and $Z_{ii}^{(k)}$.
Indeed from (\ref{dif1}) we obtain the following fundamental
relations:
\be
&& {d Z \over d x} = \tilde{\mathbb U}^{(D)}  +
(\tilde{\mathbb U}^{(O)}\ W)^{(D)} \non\\
&& {d  W\over dx } + W \tilde{\mathbb U}^{(D)} -\tilde{\mathbb
U}^{(D)}W + W(\tilde{\mathbb U}^{(O)}W)^{(D)} -\tilde{\mathbb
U}^{(O)} - (\tilde{\mathbb U}^{(O)}W)^{(O)}=0, \label{form}
\ee
where the superscripts $O,\ D$ denote off-diagonal and diagonal
part respectively. We shall come back to the solution of the latter set of equations in the subsequent sections when investigating particular physical examples.

\subsubsection{The sine-Gordon and Liouville models}
Note that we focus here on the sine-Gordon and not the sinh-Gordon model, as one would expect from the notation of the previous section, where we introduce real ATFT's. In fact the only difference compared to the notation introduced previously is that $\beta \to i \beta$; this is basically the only modification occurring between real and imaginary ATFT's.
The associated classical $r$-matrix \cite{jimbo, jimbo1} in this case is expressed as
\be
r(\lambda) = {1\over \sinh \lambda}
\begin{pmatrix}
    ({\sigma^z +1 \over 2})\cosh \lambda  & \sigma^-     \\
   \sigma^+  & ({-\sigma^z +1 \over 2})\cosh \lambda
\end{pmatrix}
\label{rjimbo}
\ee
and the Lax operator for the sine-Gordon is now given as
\be
{\mathbb U}(x,t,u) = {\beta \over 4i} \pi(x)
\sigma^z + {mu \over 4i}e^{{i \beta \over 4}\phi \sigma^z} \sigma^y e^{-{i \beta \over 4}\phi
\sigma^z}-{mu^{-1}\over 4i}e^{-{i \beta \over 4}\phi \sigma^z} \sigma^y e^{{i \beta \over 4 }\phi \sigma^z},
\label{laxsg}
\ee
$\sigma^{x, y, z}$ are the familiar 2-dimensional Pauli matrices, $u = e^{\lambda}$. ${\mathbb U}$ satisfies the fundamental quadratic algebraic relation with the classical $r$-matrix (\ref{rjimbo}), which gives rise to:
\be
\Big \{\phi(x),\ \pi (y) \Big \} = \delta(x-y).
\ee

To consider the formal series
expansion of $T$ we shall need the following symmetry of the Lax operator:
\be
{\mathbb U}(u^{-1},
\phi, \pi) = {\mathbb U} (-u. -\phi, \pi)
\ee
and consequently
\be
T(u^{-1}, \phi, \pi) = T(-u, -\phi, \pi).
\ee
As in the previous section we aim at
expressing the term of order $u$ in ${\mathbb U}$ independently of the fields, after applying a suitable gauge
transformation \cite{ftbook}. More precisely, consider the gauge transformation (\ref{decomp}) with
\be
\Omega = e^{{i\beta \over 4} \phi \sigma^z},
\ee
then the gauge transformed operator $\tilde {\mathbb U}$ is expressed as:
\be
\tilde {\mathbb U}(x,t,u)=  {\beta \over 4i} \mathfrak{f}(x) \sigma^z + {mu \over 4i} \sigma^y -{mu^{-1}\over
4i}e^{-{i \beta \over 2}\phi \sigma^z} \sigma^y e^{{i \beta \over 2 }\phi \sigma^z
}
\ee
where we define
\be
\mathfrak{f}(x,t) = \pi(x,t) + \phi'(x,t).
\ee

We consider the decomposition (\ref{exp0}) for $\tilde T$, as $|u| \to
\infty$ \cite{ftbook}.
$W,$ is off diagonal
matrix and $Z$ is purely diagonal, and are expressed as in
(\ref{expa}). Inserting  expressions (\ref{expa}) in
(\ref{newt}) one may identify the matrices $W^{(k)}$ and $Z^{(k)}$, via (\ref{form}).

It is sufficient for our purposes here to identify only the first couple of terms of
the expansions. Indeed based on equation (\ref{form}) we conclude (see also \cite{ftbook}):
\be
&& W^{(0)} = i \sigma^x, ~~~~W^{(1)} = -{i \beta \over  m} \mathfrak{f}(x) \sigma^x, \non\\
&& W^{(2)} = {2i \beta \mathfrak{f}'\over m^2}\ \sigma^y - i \sin (\beta \phi)\ \sigma^y -{\beta^2 \mathfrak{f}^2\over 2im^2}\ \sigma^x.
\ee
Then, as in the previous example we also identify the diagonal elements $Z^{(n)}$.
In particular from the first of the equations (\ref{form}) we extract the following expressions ($x=L,\ y=-L$):
\be
&&Z^{(-1)}=  -{i m L\over 2} \sigma^z, \non\\
&&Z^{(1)} = { m \over 4} \begin{pmatrix}
  -\int_{-L}^L dx\ W_{21}^{(2)}(x) &        \\
     & \int_{-L}^L dx\ W_{12}^{(2)}(x)
\end{pmatrix} - {m\over 4} \begin{pmatrix}
  -i \int_{-L}^L dx\ e^{-i \beta \phi}  &        \\
     & i \int_{-L}^L dx\ e^{i \beta \phi}
\end{pmatrix}. \non\\
\ee
Notice again that for $-i u \to \infty$ the leading contribution comes from the $Z_{11}$ elements, due to the form of $Z^{(-1)}$, see also similar argument for the NLS model described previously.

Then recalling the expression for the generating function of integrals of motion, introduced in the previous example, we conclude that
\be
I^{(1)} = Z_{11}^{(1)} = - {m \over 4i } \int_{-L}^Ldx\  \Big ( -{\beta^2 \over 2m^2}{\mathfrak f}^2 + \cos \beta \phi \Big ).
\ee
If we now do the same kind of expansion but for $\lambda \to -\infty$ we basically end up to a similar expression as before, by simply exploiting the basic symmetry of the monodromy matrix
\be
T(u^{-1}, \phi,\ \pi) = T(-u,\ -\phi,\ \pi)
\ee
we find that
\be
I^{(-1)} = - { m \over 4i } \int_{-L}^Ldx\  \Big ( -{\beta^2 \over 2m^2}\hat{\mathfrak f}^2 + \cos \beta \phi \Big ),
\ee
where we define
\be
\hat {\mathfrak f}(\phi,\ \pi) = {\mathfrak f}(-\phi,\ \pi).
\ee
In fact it is clear that any combination of the quantities $I^{(1)},\ I^{(-1)}$ is also one of the charges in involution.
So we find that the Hamiltonian is
\be
I^{(1)} + I^{(-1)} \propto {\cal H}= \int_{-L}^L dx\ \Big ( {1\over 2} (\pi^2 + \phi^{'2}) - {m^2 \over \beta^2}\cos\beta \phi\Big ),
\ee
and we may also identify the momentum as:
\be
I^{(1)} - I^{(-1)}  \propto {\cal P} = \int_{-L}^Ldx\ \phi'(x)\ \pi(x).
\ee

Let us also identify the ${\mathbb V}$ operator associated to the Hamiltonian and momentum of the sine-Gordon model.
Recall the generic expression for ${\mathbb V}$
\be
{\mathbb V}(x, \lambda, \mu,x) = t^{-1}(\lambda) tr_{a}\Big (T_a(L, x, \lambda)\ r_{ab}(\lambda -\mu)\ T_a(x, -L, \lambda) \Big ).
\ee
Recall the ansatz for $\tilde T$ (\ref{exp0}), and that the leading contribution comes from the $Z_{11}$, then the latter formula may be expressed as:
\be
{\mathbb V}(x, \lambda, \mu)  =  \Big [ (1+W_a(x))^{-1}\ \Omega^{-1}(x)\  r_{ab}(\lambda- \mu)\ \Omega(x)\  (1+W_a(x)) \Big]_{11}.
\ee
Our aim now is to expand the latter expression in powers of $u^{-1}$ (recall $u = e^{\lambda}$).
The $r$-matrix is expanded as
\be
\Omega^{-1}(x)\ r(\lambda-\mu)\ \Omega(x) = \tilde r^{(0)} + u^{-1} \tilde r^{(1)} + {\cal O}(u^{-2})
\ee
and
\be
1+W(x) & = & A + u^{-1} W^{(1)} + {\cal O}(u^{-2}) \cr
(1 + W(x))^{-1} &= & A^{-1} +  u^{-1} f^{(1)} + {\cal O}(u^{-2}),
\ee
where we define:
\be
\tilde r^{(0)} &=& {1\over 2}\begin{pmatrix}
   \sigma^z+1  & 0     \\
   0  & -\sigma^z+1
\end{pmatrix} \, ~~~~~\tilde r^{(1)} = 2v \begin{pmatrix}
   0  & e^{-{i\beta \phi \over 2}}  \sigma^-  \\
    e^{{i\beta \phi \over 2}}  \sigma^+   & 0
\end{pmatrix} \ \cr
f^{(1)} &=& {\beta \over 2m}{\mathfrak f}(x){\mathbb I}, ~~~~~A= \begin{pmatrix}
   1  & i    \\
   i & 1
\end{pmatrix} \, ~~~~A^{-1} ={1\over 2} \begin{pmatrix}
   1  & -i     \\
   -i  & 1
\end{pmatrix} \ ,
\ee
$v = e^{\mu}$.

We keep for our purposes here only the first terms of the expansion; these are sufficient to provide the relevant ${\mathbb V}$
quantities.
Gathering all the information from the latter expression we eventually obtain:
\be
{\mathbb V}^{(1)}(x, \mu) ={\beta \over 2m} {\mathfrak f}(x, t) \sigma^z + v \Omega(x, t) \sigma^y \Omega^{-1}(x,t).
\ee
Similar computation for the expansion in powers of $u$ can be made and the final result in his case is directly given as:
\be
{\mathbb V}^{(-1)}(x, \mu) =- {\beta \over 2m} \hat {\mathfrak f}(x, t) \sigma^z + v^{-1} \Omega^{-1}(x, t) \sigma^y \Omega(x,t).
\ee
By adding the two expressions above we end to the desired result (we have also multiplied with an overall factor ${m\over 4i}$)
\be
{\mathbb V}^{({\cal H})} = {\beta \over 4 i}\phi'(x, t)\sigma^z + {m v \over 4i} \Omega(x, t)\sigma^y \Omega^{-1}(x,t) + {mv^{-1} \over 4i} \Omega^{-1}(x,t) \sigma^y \Omega(x, t).
\ee
Similarly, subtraction of the two quantities ${\mathbb V}^{(1)},\ {\mathbb V}^{(-1)}$ leads to the time component of the Lax pair associated to the momentum, which coincides with ${\mathbb U}$.

Note that appropriate massless limit of the sine-Gordon model gives rise to the
Liouville theory through a particular limiting process (see e.g. \cite{fati, doikouxxz}).
The entailed Lax pair of the Liouville theory takes the form:
\be
{\mathbb U}(\lambda)= {1\over 2} \left(
\begin{array}{cc}
-i\pi(x)   & -2 e^{-\lambda -i\phi(x)} \\
4 \sinh(\lambda -i\phi(x)) &i\pi(x)\\
\end{array}\right ), ~~~~{\mathbb V}(\lambda)= {1\over 2} \left(
\begin{array}{cc}
-i\phi'(x)   & 2 e^{-\lambda -i\phi(x)} \\
4\cosh(\lambda -i\phi(x)) &i\phi'(x)\\
\end{array}\right ).  \label{liouc}
\ee
The Lax pair satisfies the zero curvature condition,
which leads to the corresponding
equations of motion i.e.
\be
&& \mbox{sine-Gordon model:}
~~~~~\ddot{\phi}(x) - \phi''(x) +{m^{2} \over \beta}
\sin(\beta \phi(x))=0 \non\\
&& \mbox{Liouville model:}
~~~~~\ddot{\phi}(x) - \phi''(x) -4 i e^{-2i\phi(x)}=0.
\ee
\\
\textbf{\underline{Exercise 7.2.}} Starting from the ${\mathbb U}$ operator of the
Liouville model derive the energy and momentum of the model well as the
associated ${\mathbb V}$ operators. Determine also the corresponding equations of motion.

\subsubsection{The $A_2^{(1)}$ ATFT}
We shall now study the second member of the ATFT hierarchy, i.e. the $A_2^{(1)}$ model.
It will be useful in what follows to introduce some notation (see also \cite{doikouatft, doikouatft1}):
\be
{\beta \over 2} \Theta \cdot H = \mbox{diag} (a,\ b,\ c),
~~~~~{\beta \over 2} \hat \Theta \cdot H = \mbox{diag} (\hat a,\
\hat b,\ \hat c), ~~~~ e^{\beta \alpha_i \cdot \Phi} = \gamma_i
\label{def}
\ee
$\hat \Theta(\Phi,\ \Pi) = \Theta(-\Phi,\ \Pi)$ explicit expression of $a,\ b,\ c$ and $\gamma_i$
can be found in Appendix A.
From the first of equations (\ref{form}) we may derive the
matrices $Z,\ \hat Z$. Indeed one may easily show that:
\be
{d Z^{(0)}\over dx } &=& {m\over 4} \left(
\begin{array}{ccc}
 W_{21}^{(1)} + \zeta a     &  & \\
                    & W_{32}^{(1)} + \zeta b &  \\
 &  &  -W_{13}^{(1)} + \zeta c\\ \end{array} \right) =0 \non\\
{d \hat Z^{(0)}\over dx }   &=& {m\over 4} \left(
\begin{array}{ccc}
-\hat W_{31}^{(1)}  +\zeta \hat a    &  & \\
                                          & \hat W_{12}^{(1)}  + \zeta \hat b &  \\
 &  & \hat W_{23}^{(1)}  +\zeta \hat c\\ \end{array} \right)=0.
\ee
It is clear that the latter quantities are zero because of the
form of $W_{ij}^{(1)},\ \hat W_{ij}^{(1)}$ see Appendix A. Also
the higher order $Z^{(k)},\ \hat Z^{(k)}$ are given by:
\be {d
Z^{(k)}\over dx } &=& {m\over 4} \left(
\begin{array}{ccc}
W_{21}^{(k+1)}  -\gamma_3 W_{31}^{(k-1)}     &  & \\
                                          & W_{32}^{(k+1)}  +\gamma_1 W_{12}^{(k-1)} &  \\
 &  & -W_{13}^{(k+1)}  +\gamma_2 W_{23}^{(k-1)}\\ \end{array} \right) \non\\
{d \hat Z^{(k)}\over dx }   &=& {m\over 4} \left(
\begin{array}{ccc}
-\hat W_{31}^{(k+1)}  +\gamma_1 \hat W_{21}^{(k-1)}     &  & \\
                                          & \hat W_{12}^{(k+1)}  +\gamma_2 \hat W_{32}^{(k-1)} &  \\
 &  & \hat W_{23}^{(k+1)}  -\gamma_3 \hat  W_{13}^{(k-1)}\\ \end{array} \right)
\non\\ k > 0. \label{zz0}
\ee
The computation of $W,\ \hat W$ is
essential for defining the diagonal elements. First it is
important to discuss the leading contribution of the above
quantities as $|u| \to \infty$. To achieve this we shall need the
explicit form of $Z^{(-1)},\ \hat Z^{(-1)}$:
\be
Z^{(-1)}(x, y) =
{m (x-y)\over 4} \left( \begin{array}{ccc}
e^{{i \pi \over 3}}     &  & \\
                                          &e^{-{i \pi \over 3}}  &  \\
 &  & -1 \\ \end{array} \right), ~~~\hat Z^{(-1)}(x,y)=  {m (x-y)\over 4}
\left( \begin{array}{ccc}
e^{-{i \pi \over 3}}     &  & \\
                                          &e^{{i \pi \over 3}}  &  \\
 &  & -1 \\ \end{array} \right). \non\\ \label{zz}
\ee
The information above will be extensively used subsequently.
From the formulas (\ref{form}), (\ref{zz0}) the matrices
$W^{(k)},\ \hat W^{(k)},\ Z^{(k)},\ \hat Z^{(k)}$ may be
determined (see Appendix for details).

The local local
integrals of motion in the periodic case, emerge as usually from the
expansion ($|u|\to \infty$) of:
\be \ln\ [tr T(u)] = \ln\ \Big [tr
\Big\{(1+W(L,u))\ e^{Z(L, -L, u)}\ (1+W(-L,u))^{-1}\Big \}\Big].
\ee
Notice
that in the case of periodic boundary conditions we put our system
in the `whole' line ($x = L,\ y =-L$), and consider Schwartz
boundary conditions, i.e. the fields and their derivatives vanish
at the end points $\pm L$. Bearing in mind that as $u \to -\infty$
the leading contribution of $e^{Z},\ (e^{\hat Z})$ (see
(\ref{zz})) comes from the $e^{Z_{33}},\ (e^{\hat Z_{33}})$ term,
the expression above becomes
\be
\ln\ [tr T(u \to -\infty)] =
\sum_{k= -1} {Z^{(k)}_{33} \over u^k}.
\ee
To reproduce the
familiar local integrals of motion we shall need both $Z(L, -L,
u),\ \hat Z(L, -L, u)$. Let
\be
&& I^{(1)} = -{12 m \over
\beta^{2} } Z_{33}^{(1)}(L, -L, u) = \int_{-L}^{L} dx \Big(
\sum_{i=1}^2 \theta_i^2  + {m^2 \over \beta^2}
\sum_{i=1}^3 e^{\beta \alpha_i \cdot \Phi}  \Big ), \non\\
&& {I}^{(-1)} = -{12 m \over \beta^{2} }
\hat Z_{33}^{(1)}(L, -L, u) = \int_{-L}^{L} dx
\Big( \sum_{i=1}^2 \hat \theta_i^2 + {m^2 \over \beta^2}
\sum_{i=1}^3 e^{\beta \alpha_i \cdot \Phi}  \Big )
\ee
the momentum and
Hamiltonian (and the higher conserved quantities) of the ATFT are
given by:
\be
&& {\cal P}_{1} = {1\over 2}(I^{(-1)} - I^{(1)} ) = \int_{-L}^{L} dx
\sum_{i=1}^2 \Big ( \pi_i\ \phi_i' - \pi_i'\ \phi_i \Big ) \non\\
&& {\cal H}_{1} = {1\over 2} (I^{(1)} + I^{(-1)}) =
\int_{-L}^{L} dx \Big( \sum_{i=1}^2 (\pi_i^2 + \phi_i^{'2}) + {m^2
\over \beta^2}
\sum_{i=1}^3 e^{\beta \alpha_i \cdot \Phi}  \Big ). \non\\ \label{im}
\ee
Note that the boundary terms are absent in the expressions above, since we considered
Schwartz type boundary conditions at $\pm L$. Also, in the generic situation,
for any $A_{n}^{(1)}$, the sum in the momentum ${\cal P}_1$ and the
kinetic term of the Hamiltonian ${\cal H}_1$ runs from 1 to $n$,
whereas the sum in the potential term of the Hamiltonian runs from 1
to $n+1$. A study of generic integrable boundary conditions in ATFT's
is presented in \cite{corrigan1}--\cite{doikouatft1}.
\\
\\
\underline{\textbf{Exercise 7.2.}}
Based on the process described for the derivation of Lax pairs,
determine the ${\mathbb V}$-operator associated to the integral of motion ${\cal H}_1$ (\ref{im}).

\section{The continuum integrable limit}
The main aim in this section is to introduce a consistent continuum
limit of integrable discrete models in such a way that
integrability is preserved (see also \cite{ads}).
The starting point is the lattice monodromy matrix, which is expressed as:
\be
T_a = L_{a N}\ L_{a N -1} \ldots L_{a 1}.
 \label{lwlL2}
\ee
Assume that $L$ admits an expansion in powers of $\delta$ as
\be
L_{ai}= 1 + \delta {\mathbb U}_{ai} + {\cal O}(\delta^2)\ ,
\ee
then consider the product:
\be
T_a = \prod_{i=1}^N (1 + \delta {\mathbb U}_{ai} + \sum_{n=2}^{\infty} \delta^n {\mathbb U}^{(n)}_{ai}))\ .
\ee
Expanding the expression above in powers of $\d$, we get
\be
T_a = 1 + \delta \sum_i {\mathbb U}_{ai} + \delta^2\sum_{i<j} {\mathbb U}_{a_{i}} {\mathbb U}_{aj} + \delta^2 \sum_{i} {\mathbb U}^{(2)}_{ai}
+\dots \ .
\ee
These, multiple in general, infinite series of the products of local terms,
are characterized by two indices: the overall power $n$ of $\delta$, and the number $m$
of the set of indices $i$
over which the series is summed. Note that, in the $T$ expansion one always has $n\geqslant m$.
The continuum limit soon to be defined more precisely, will entail the limit $\delta \to 0$ with
${\cal O}(N) = {\cal O}({1/\delta})$.
We now formulate the following {\it power-counting rule}, that is terms of the form (see also \cite{ads})
\be
\delta^n \sum_{i_1< i_2 <\dots i_m}{\mathbb U}_{ai_1}^{(n_1)} ...{\mathbb U}_{ai_m}^{(n_m)}\ ,
\qq \sum_{j=1}^m n_j =n\ ,
\ee
with $n>m$ are omitted in the continuum limit. The latter is defined by
\be
\delta \sum_{i} {\mathbb U}_{ai} \to \int_{-L}^{L} dx\ {\mathbb U}_a(x)\
\ee
and similarly for multiple integrals. Here $2L$ is the length of the continuous interval
defined as the limit of $N \delta$.
In other words, contributions to the continuum limit may only come from the terms with $n=m$
for which the power $\delta^n$ can be exactly matched by the ``scale'' factor $N^m$ of the $m$-multiple sum over
$m$ indices $i$. In particular, only terms of order one
in the $\delta$ expansion of local classical matrices $L_{ai}$ will contribute to the continuum limit.
Any other contribution acquires a scale factor $\delta^{n-m} \to 0$, when the continuum limit is taken (see also \cite{ads}).
This argument is of course valid term
by term in the double expansion, and it always has to be checked for consistency.

The continuous limit
of $T$, hereafter denoted ${\cal T}$,
is then immediately identified as the path-ordered
exponential from $y=-L$ to $x=L$
\be
{\cal T} = P \exp{\left (\int_{-L}^{L} dx\  {\mathbb U}(x) \right )}\ ,
\ee
where suitable (quasi) periodicity conditions on the
continuous variables of the classical matrix ${\mathbb U}(x)$ are assumed.
Of course the derivation of a continuous limit requires that the $L$-matrices are not
\cal{too} inhomogeneous (e.g. $L$-matrices at neighbor sites should not be too different.

The above identification of ${\cal T}$ has been built so that to straightforwardly
generate the classical continuous limit of the Hamiltonians from the analytic expansion
\be
{\rm tr} ({\cal T}(\lambda)) \equiv \sum_{n=1}^{\infty} (\lambda - \lambda_{0})^n {\cal H}^{(n)} \ .
\label{fhhjh2}
\ee
We thus characterize ${\mathbb U}(x)$ as a local Lax matrix yielding
the hierarchy of continuous Hamiltonians ${\cal H}^{(n)}$. In order for this statement to agree with the
key assumption of preservation of integrability we are now lead to
require a Poisson structure for ${\mathbb U}$
compatible with the demand of classical integrability of the continuous Hamiltonians.
Indeed, such a structure is deduced as the ultra-local Poisson bracket
\be
\Big \{{\mathbb U}_1(x, \lambda_1),\ {\mathbb U}_2(y, \lambda_2)\Big \}
= \Big [r_{12}(\lambda_1 -\lambda_2),\ {\mathbb U}_1(x, \lambda_1)+{\mathbb U}_2(y, \lambda_2)\Big ] \delta(x-y)\ ,
\label{funda1}
\ee
where $r$ is the classical matrix characterizing the exchange algebra
of the $L$-operators.
More specifically, recalling that $L_{ai} = 1 + \delta {\mathbb U}_{ai} + {\cal O}(\delta^2)$, plugging it into (\ref{clalg})
and assuming ultra-locality of Poisson brackets one gets
\be
\Big \{{\mathbb U}_{ai}(\lambda_1),\ {\mathbb U}_{bj}(\lambda_2)\Big \} = \Big [r_{ab}(\lambda_1-\lambda_2),\ {\mathbb U}_{ai}(\lambda_1) + {\mathbb U}_{bj}(\lambda_2) \Big ]{\delta_{ij} \over \delta}\ .
\ee
One then identifies,  in the continuum limit $\delta \to 0$, the factor
 ${\delta_{ij}/ \delta}$ with $\delta(x-y)$.
Reciprocally, it is a well known result (see, for instance \cite{ftbook}) that if ${\mathbb U}(x)$ has a such an ultra-local linear Poisson bracket (\ref{funda1}) the full
monodromy matrix between $-L$ and $L$ has the quadratic Poisson bracket structure, thereby guaranteeing
Poisson commutation of the Hamiltonians.

We now focus on the continuum limit of the zero curvature condition, and recover the continuum expression as a suitable continuum limit of the discrete one. Moreover, the generic continuum expression for the ${\mathbb V}$-operator is also ensued via this process. Recall  the discrete zero curvature condition:
\be
\dot{L}_n(\lambda) = {\mathbb A}_{n+1}(\lambda)L_n(\lambda)- L_n(\lambda) {\mathbb A}_n(\lambda).
\ee
Consider the following identifications:
\be
L_n \to 1 +\delta {\mathbb U}(x), ~~~~~{\mathbb A}_n \to {\mathbb V}(x), ~~~~~{\mathbb A}_{n+1} \to {\mathbb V}(x+\delta)
\ee
then expressing
\be
{\mathbb V}(x+\delta) = {\mathbb V}(x) + \delta {\mathbb V}'(x) +{\cal O}(\delta^2)
\ee
and keeping the first non trivial contribution which is of order $\delta$ we conclude:
\be
\dot{\mathbb U}(\lambda, x) - {\mathbb V}'(\lambda, x) + \Big [{\mathbb U}(\lambda,x),\ {\mathbb V}(\lambda, x)\Big ] =0,
\ee
which is nothing else but the familiar continuum zero curvature condition.
\\
\\
\textbf{\underline{Example.}} Let us work out a particular example to demonstrate how this process may be applied.
Consider the discrete NLS model; first introduce the spacing parameter $\delta$ in the Lax operator (see also \cite{doikoudefect} and references therein):
\be
L(\lambda) = \begin{pmatrix}
   \delta \lambda +1 - \delta^2{\mathrm x}{\mathrm X}  &\delta {\mathrm x} \\
    -\delta {\mathrm X} & 1
\end{pmatrix} \ \label{laxd}
\ee
we have essentially rescaled the fields ${\mathrm x},\ {\mathrm X}$ and the spectral parameter as:
\be
\lambda \to \delta \lambda, ~~~~~{\mathrm x }\to \delta {\mathrm x}, ~~~~~{\mathrm X} \to \delta {\mathrm X}. \label{id1}
\ee
After the aforementioned rescaling the first local integrals of motion become:
\be
&&  I_1 = \sum_j(1 -\delta^2 {\mathrm x}_j {\mathrm X}_j) \non\\
&& I_2 = -\delta^2 \sum_{j=1}^{N-1} {\mathrm x}_{j+1}{\mathrm X}_j - {1\over 2} \sum_{j=1}^N(1 - \delta^2 {\mathrm x}_j
{\mathrm X}_j)^2.
\ee
We shall consider the continuum limit of the first two integrals just to illustrate how this works. Bear in mind the following identifications:
\be
&& {\mathrm x}_j \to {\mathrm x}(x), ~~~~~{\mathrm X}_j \to {\mathrm X}(x) \non\\
&& {\mathrm x}_{j+1} \to {\mathrm x}(x+\delta), ~~~~~~{\mathrm X}_{j+1} \to {\mathrm X}(x+\delta), \label{id2}
\ee
also consider the approximation for the sum
\be
\delta \sum_{j=1}^N f_j \to \int_{-L}^L dx\ f(x).
\ee
It is then straightforward to show that the first integral of motion, after keeping the first non trivial contributions of order $\delta$, becomes in the continuum limit:
\be
{\cal I}_1 = \int_{-L}^L {\mathrm x}(x) {\mathrm X}(x).
\ee
Similarly, the second integral becomes after taking the continuum limit and keeping the next nontrivial order $\delta^2$,
\be
{\cal I}_2 = {1\over 2} \int_{-L}^L dx\ \Big ( {\mathrm x}'(x){\mathrm X}(x) - {\mathrm x}(x){\mathrm X}'(x) \Big ).
\ee
It is also clear that the continuum limit of the associated Lax operator (\ref{laxd}) takes the desired form:
\be
L(\lambda) \to 1 + \delta {\mathbb U}(x) +{\cal O}(\delta^2),
\ee
where we define
\be
{\mathbb U}(x) = \begin{pmatrix}
   {\lambda \over 2} &{\mathrm x}(x) \\
 -{\mathrm X}(x)     & -{\lambda \over 2}
\end{pmatrix},
\ee
notice that ${\mathbb U}$ may be always derived up to an additive constant.
Take also the continuum limit of the ${\mathbb A}$ operator associated to the second charge ${\cal H}_2$
this then becomes via (\ref{id1}), (\ref{id2})
\be
{\mathbb A}_n \to {\mathbb V}(x), ~~~~~{\mathbb V}(x) = \begin{pmatrix}
   \lambda   &{\mathrm x}(x) \\
    -{\mathrm X}(x) & 0
\end{pmatrix}.
\ee

By identifying
\be
{\mathrm x}(x) \equiv \bar \psi(x), ~~~~~{\mathrm X}(x) \equiv - \psi(x)
\ee
we end up to the familiar continuum expressions extracted in the previous section. In particular, ${\cal I}_1$ and ${\cal I}_2$
are proportional to ${\cal N}$ and ${\cal P}$ respectively. This is a strong consistency check verifying the validity of the applied continuum limit process.
\\
\\
\underline{\textbf{Exercise 8.1.}}
Determine the continuum limit of the discrete NLS Hamiltonian ${\cal I}_3$.

\appendix

\section{$W$ and $Z$ matrices for the $A_2^{(1)}$ ATFT}

We write below explicit expressions of
the $W^{(n)},\ Z^{(n)}$ matrices for the first orders:
\be
&& W^{(0)} = \hat
W^{(0)}= \left(
\begin{array}{ccc}
0                    & e^{{i \pi \over 3}}  & 1 \\
e^{{i \pi \over 3}}  & 0                    & -1 \\
e^{{2i \pi \over 3}} & e^{-{i \pi \over 3}} & 0\\ \end{array} \right), \non\\
&& {m\over 4} W^{(1)}= \left( \begin{array}{ccc}
0       & e^{{2i \pi \over 3}} a  & c \\
-a      & 0                       & b \\
e^{{i \pi \over 3}}c  & -b        & 0\\ \end{array} \right), ~~~~{m\over 4}
\hat W^{(1)} = \left( \begin{array}{ccc}
0       & -\hat b  & -\hat a \\
-e^{-{i \pi \over 3}}\hat b      & 0                       & -\hat c \\
\hat a  & -e^{{i \pi \over 3}}\hat c        & 0\\ \end{array}
\right).
\ee
The higher order quantities are more complicated and
we give the matrix entries below for $W^{(2)},\ \hat W^{(2)}$
(define also, $~\zeta = {4 \over m}$):
\be
&& W_{12}^{(2)} =
{1\over 3} (-2 \gamma_3 + \gamma_1 +\gamma_2) + {\zeta^2 \over 3}
(2a' + b') + {\zeta^2 \over 3}
(-2a^2 -bc), \non\\
&& W^{(2)}_{21}= {e^{-{i\pi \over 3}} \over 3} (-2\gamma_3 + \gamma_1 +\gamma_2)
 + {\zeta^2 e^{-{i\pi \over 3}} \over 3} (a' - c') + {\zeta^2 e^{-{i\pi \over 3}}
\over 3}(c^2 -ab) \non\\
&& W_{13}^{(2)} =  {1\over 3}
(-2 \gamma_2+ \gamma_1 +\gamma_3) + {\zeta^2 \over 3} (-b' + c') +
{\zeta^2 \over 3}(b^2 -ac), \non\\
&& W_{31}^{(2)} =  {1\over 3}  (2 \gamma_2 - \gamma_1 -\gamma_3) +
{\zeta^2 \over 3} (-a' -2c') + {\zeta^2 \over 3}(2c^2 + a b), \non\\
&& W_{23}^{(2)} =  -{1\over 3}
(2 \gamma_1 -\gamma_2 - \gamma_3 ) + {\zeta^2 \over 3} (2b' + c') +
{\zeta^2 \over 3}(-2b^2 -ac) \non\\
&& W^{(2)}_{32}= -{e^{{i\pi \over 3}} \over 3}
(2 \gamma_1 - \gamma_2 - \gamma_3)
 + {\zeta^2 e^{{i\pi \over 3}} \over 3} (-a' +b') +
{\zeta^2 e^{{i\pi \over 3}} \over 3}(a^2 -b c) \ee and \be && \hat
W_{12}^{(2)} =  {e^{-{i\pi \over 3}} \over 3} (-2 \gamma_2 +\gamma_1
+ \gamma_3) + {\zeta^2e^{-{i\pi \over 3}} \over 3} (\hat  b' - \hat
c') + {\zeta^2 e^{-{i\pi \over 3}}
\over 3}(\hat c^2 - \hat a \hat b), \non\\
&& \hat W^{(2)}_{21}= {1 \over 3} (-2 \gamma_2 + \gamma_1 +\gamma_3)
 + {\zeta^2  \over 3} (2\hat b' +\hat a')+ {\zeta^2 \over 3}
(-2 \hat b^2 -\hat a \hat c) \non\\
&& \hat W_{13}^{(2)} =  -{1\over 3} (-2 \gamma_1  +\gamma_3 +\gamma_2) - {\zeta^2 \over 3}
(2\hat a' + \hat c') +
{\zeta^2 \over 3} (2\hat a^2 +\hat b \hat c), \non\\
&& W_{31}^{(2)} =  {e^{{i\pi \over 3}}\over 3}  (2 \gamma_1 -\gamma_2-\gamma_3) +
{\zeta^2 e^{{i\pi \over 3}} \over 3} (\hat b' -\hat a') + {\zeta^2 e^{{i\pi \over 3}}
\over 3}(-\hat b^2 +\hat a \hat c), \non\\
&& \hat W_{23}^{(2)} =  -{1\over 3}
(-2 \gamma_3  +\gamma_2 + \gamma_1 ) + {\zeta^2 \over 3} (\hat a' -\hat c') + {\zeta^2 \over 3}
(-\hat a^2 + \hat b \hat c) \non\\
&& \hat W^{(2)}_{32}= {1 \over 3} (-2 \gamma_3 +\gamma_1 + \gamma_2 )
 + {\zeta^2  \over 3} (\hat b' + 2 \hat c') + {\zeta^2  \over 3}(-2 \hat c^2 -
\hat a \hat b) \ee where the prime denotes derivative with respect
to $x$, also $a,\ b,\ c,$ and $\gamma_i$ are defined in
(\ref{def}) and have the following explicit forms: \be && a =
{\beta \over 2} ({\theta_1 \over 2} + {\theta_2 \over 2 \sqrt 3}
), ~~~~b = {\beta \over 2} (-{\theta _1 \over 2}+ {\theta_2 \over
2 \sqrt 3}),
~~~~c = -{\beta \over 2}  {\theta_2 \over  \sqrt 3}, \non\\
&& \gamma_1 = e^{\beta \phi_1}, ~~~~\gamma_2 = e^{\beta(-{1 \over 2}\phi_1 + {\sqrt 3 \over 2} \phi_2)},
 ~~~~~\gamma_3=e^{\beta(-{1 \over 2}\phi_1 - {\sqrt 3 \over 2} \phi_2)}. \label{def2}
\ee
Moreover using the expressions above and (\ref{zz0}) we have:
\be
&& {d Z_{11}^{(1)}\over dx} = {e^{-{i\pi \over 3}}\over 3} {m
\over 4}(\gamma_1 +\gamma_2 + \gamma_3)  +{\zeta e^{-{i\pi \over
3}}\over 3} (a' -c') +{\zeta e^{-{i\pi \over 3}} \over 6} (a^2
+b^2 +c^2) \non\\ && {d Z_{22}^{(1)} \over dx} = {e^{{i\pi \over
3}} \over 3} {m \over 4}(\gamma_1 +\gamma_2 + \gamma_3)  +{\zeta
e^{{i\pi \over 3}} \over 3} (b' -a') +{\zeta e^{{i\pi \over
3}}\over 6} (a^2 +b^2 +c^2) \non\\ && { d Z_{33}^{(1)} \over dx} =
-{1\over 3} {m \over 4}(\gamma_1 +\gamma_2 + \gamma_3)  -{\zeta
\over 3} (c' -b') -{\zeta \over 6}
(a^2 +b^2 +c^2) \non\\
&& {d \hat Z_{11}^{(1)} \over dx} = {e^{{i\pi \over 3}} \over 3}
{m \over 4}(\gamma_1 +\gamma_2 + \gamma_3) -{\zeta e^{{i\pi \over
3}} \over 3} (\hat b' -\hat a') +{\zeta e^{{i\pi \over 3}}\over 6}
(\hat a^2 +\hat b^2 +\hat c^2)
\non\\
&& {d \hat Z_{22}^{(1)} \over dx} = {e^{-{i\pi \over 3}}\over 3}
{m \over 4}(\gamma_1 +\gamma_2 + \gamma_3)  +{\zeta e^{-{i\pi
\over 3}}\over 3} (\hat b' - \hat c') +{\zeta e^{-{i\pi \over 3}}
\over 6} (\hat a^2 +\hat b^2 +\hat c^2) \non\\ && {d \hat
Z_{33}^{(1)}\over dx} = -{1\over 3} {m \over 4}(\gamma_1 +\gamma_2
+ \gamma_3) +{\zeta \over 3} (\hat a' -\hat c') -{\zeta \over 6}
(\hat a^2 +\hat b^2 + \hat c^2). \label{z1}
\ee
\\
\\
{\bf Acknowledgments}
\\
I wish to thank the Department of Physics of the University of Oldenburg, and in particular J. Kunz for kind hospitality.
I am also indebted to J. Avan for useful suggestions and comments on the manuscript.
\\

\end{document}